\documentclass[a4paper,aps,pre,superscriptaddress,floatfix,nofootinbib,onecolumn,longbibliography]{revtex4}
\usepackage[pdftex]{graphicx}
\usepackage{latexsym,amsmath,verbatim,amssymb,txfonts}
\usepackage{lipsum}
\usepackage{color}
\usepackage{rotating}
\usepackage{verbatim}
\usepackage{multirow}
\usepackage[english]{babel}
\usepackage{comment}
\usepackage{bm}
\usepackage{amsmath}
\usepackage{amsfonts}
\usepackage{amssymb}
\usepackage{float}
\usepackage{color}
\usepackage{braket}
\usepackage{blkarray, bigstrut}
\usepackage{subfigure}
\usepackage{comment}

\usepackage{mathtools}
\usepackage{graphicx}
\usepackage{tikz}
\usepackage{xr}
\newtheorem{theorem}{Theorem}[section]
\newtheorem{lemma}[theorem]{Lemma}

\definecolor{internationalkleinblue}{rgb}{0.0, 0.18, 0.65}
\definecolor{brickred}{rgb}{0.8, 0.25, 0.33}
\definecolor{amber}{rgb}{1.0, 0.75, 0.0}
\definecolor{applegreen}{rgb}{0.55, 0.71, 0.0}
\definecolor{darkviolet}{rgb}{0.58, 0.0, 0.83}

\newcommand{\cinzia}[1]{\textcolor{cyan}{\textbf{#1}}}

\begin{document}

\title{Taming Cluster Synchronization}

\author{Cinzia Tomaselli}
\affiliation{Department of Electrical, Electronics and Computer Science Engineering, University of Catania, Italy}

\author{Lucia Valentina Gambuzza}
\affiliation{Department of Electrical, Electronics and Computer Science Engineering, University of Catania, Italy}

\author{Gui-Quan Sun}\email{Corresponding author: gquansun@126.com}
\affiliation{Sino-Europe Complex Science Center, North University of China, Shanxi, Taiyuan 030051, China}
\affiliation{Complex Systems Research Center, Shanxi University, Shanxi, Taiyuan 030006, China}

\author{Stefano Boccaletti}
\affiliation{Institute of Interdisciplinary Intelligent Science, Ningbo Univ. of Technology, Ningbo, China}
\affiliation{Sino-Europe Complex Science Center, North University of China, Shanxi, Taiyuan 030051, China}
\affiliation{CNR - Institute of Complex Systems, I-50019 Sesto Fiorentino, Italy}

\author{Mattia Frasca}\email{Corresponding author: mattia.frasca@unict.it}
\affiliation{Department of Electrical, Electronics and Computer Science Engineering, University of Catania, Italy}

\date{\today}

\begin{abstract}

Synchronization is a widespread phenomenon observed across natural and artificial networked systems. It often manifests itself by clusters of units exhibiting coincident dynamics. These clusters are a direct consequence of the organization of the Laplacian matrix eigenvalues into spectral localized blocks. We show how the concept of spectral blocks can be leveraged to design straightforward yet powerful controllers able to fully manipulate cluster synchronization of a generic network, thus shaping at will its parallel functioning. Specifically, we demonstrate how to induce the formation of spectral blocks in networks where such structures would not exist, and how to achieve precise mastering over the synchronizability of individual clusters by dictating the sequence in which each of them enters or exits the synchronization stability region as the coupling strength varies. Our results underscore the pivotal role of cluster synchronization control in shaping the parallel operation of networked systems, thereby enhancing their efficiency and adaptability across diverse applications.

\end{abstract}

\maketitle

From brain dynamics and neuronal firing to epidemics and power grids, synchronization plays a pivotal role in determining the collective behavior of networked dynamical units \cite{pikovskibook,boccaphysrep,boccabook}.  In fact, the assumption that the whole network synchronizes uniformly oversimplifies what actually happens in many real-world systems, where regular functioning often relies, instead, on cluster synchronization (CS). CS is a state wherein specific clusters of nodes within the network synchronize internally while exhibiting distinct dynamics from other clusters ~\cite{sorrentino2007,nicosia2013,ji2013,williams2013,pecora2014,sorrentino2016}. The phenomenon is particularly relevant in systems where parallel processing or localized information exchange is essential, such as brain networks and communication networks. Gathering a full control of CS is therefore of crucial value for the optimization of the performance of such systems.

Despite the recognized importance of CS, the development of control methods tailored for taming and regulating clustered states is currently lagging behind, if compared to techniques for controlling global synchronization. Several control strategies exist, indeed, for promoting or suppressing synchronization in entire networks \cite{porfiri2005,liu2011,zhang2013, YuDeLellisChenBernardo2012,RosenblumArkady2004, YuChenJinhu2009, Zhang2007} or in groups of symmetrical nodes \cite{Gambuzza2018,Lin2016,Fu2013,gambuzza2020controlling}, but their extension to a generic cluster-level dynamics is still unavailable. The heterogeneous nature of clusters, together with the intricate structure of inter-cluster interactions, calls therefore for novel control methods capable of selectively targeting and manipulating specific splay states within the network. Addressing this gap in knowledge, as we do in our work, represents a pressing frontier in the field of network science and control theory, and offers unprecedented opportunities for unlocking the full potential of complex networked systems across diverse applications.

We, therefore, start by considering a pristine network $\mathcal{G}$ of $N$ identical dynamical systems described by
\begin{equation}
    \dot{\mathbf{x}}_i=\mathbf{f}(\mathbf{x}_i)+d \sum_{j=1}^N a_{ij}\left(\mathbf{h}(\mathbf{x}_j)-\mathbf{h}(\mathbf{x}_i)\right)+\mathbf{u}_i
    \label{eq_main: system_adjacency}
\end{equation}
\noindent with $i=1, \ldots, N$. Here, $\mathbf{f}$ is a generic flow governing the uncoupled dynamics of each unit, $d$ is a coupling strength, $a_{ij}$ are the entries of the adjacency matrix $\mathrm{A}=\{a_{ij}\}$ ($a_{ij}=1$ if nodes $i$ and $j$ are connected and $a_{ij}=0$ otherwise), $\mathbf{h}$ is a (again generic) coupling function, and $\mathbf{u}_i$ a control input. The only assumption made, for the time being, is that the interaction graph associated with $\mathrm{A}$ is undirected and weighted.

We focus on the synchronous behavior of the network and, in particular, on the onset of clusters of synchronous nodes. Specifically, we want to describe a scenario where, given a set of $M$ network's clusters (denoted by $C_1, C_2, \ldots, C_M$), one or more of them display a synchronous dynamics i.e., $\lim\limits_{t \rightarrow +\infty} \| \mathbf{x}_i(t)-\mathbf{x}_j(t)\|=0$ $\forall i,j \in C_l$ for some $l \in \{1,\ldots, M\}$. Ref. \cite{bayani2023transition} has proved rigorously that such clustered states correspond to the presence of spectral blocks in the structure of $\mathcal{G}$.
A spectral block $\mathcal{S}$ localized at nodes $\{i_1, i_2, \ldots, i_{N'}\}$ is defined as a subset of $(N'-1)$ eigenvectors of the Laplacian matrix ${\cal L}$ associated to $\mathcal{G}$ displaying the following features:  {\it i) } all $\mathbf{v} \in \mathcal{S}$ [$\mathbf{v} \equiv (v_1, v_2, ... , v_N)$ ] are such that $v_i=0$ $\forall i \notin {i_1, i_2, \ldots, i_N'}$, and  {\it ii) } all $\mathbf{v} \notin \mathcal{S}$ are such that $v_i=v_j$ $\forall i,j \in {i_1, i_2, \ldots, i_N'}$. Ref. \cite{bayani2023transition} demonstrated that a group of nodes $C$ is associated with a spectral block $\mathcal{S}$ {\it if and only if} they are equally connected (i.e., with the same weight) to all other nodes of the network.

Nodes associated with spectral blocks therefore receive a same input from the rest of the network, and as so they form a cluster that can synchronize {\it independently} on the dynamics of all the other nodes. The stability of the synchronous clustered states associated with spectral blocks can be assessed with good approximation by using an approach based on the Master Stability Function (MSF, see Ref. \cite{pecora1998master}, and our Supplementary Material). In particular, here we focus on the challenging case of a type III MSF \cite{bibbia}. In that case, denoting by $\mathcal{L}'$  the Laplacian matrix of the subgraph $\mathcal{G'}$ associated to the cluster $C$ and calling $s$ the strength through which each node in $C$ is connected with the rest of the graph,  the condition for synchronization is given by $d\lambda_i(\mathcal{L}') \in [\nu_1^*, \nu_2^*]$ $\forall i =2, \ldots, N'$, and $\frac{\lambda_{N'}(\mathcal{L}')+s}{\lambda_2(\mathcal{L}')+s} < \frac{\nu_2^*}{\nu_1^*}$. Here $\nu_1^*$ and $\nu_2^*$ are the two critical values  at which the MSF $\lambda_{\rm max}(\nu)$  crosses the x-axis, namely  $\lambda_{\rm max}(\nu)<0$ for $\nu \in [\nu_1^*, \nu_2^*]$ (see SM for full details).

The control input $\mathbf{u}_i$ in Eq. (\ref{eq_main: system_adjacency}) is written as
\begin{equation}
    \mathbf{u}_i= \sum_{j=1}^N{w'}_{ij}\left(\mathbf{h}(\mathbf{x}_j)-\mathbf{h}(\mathbf{x}_i)\right),
    \label{eq_main: system_adjacency_controlw'}
\end{equation}

\noindent where $w'_{ij}$ are the weights of the control links added to the pristine network. We move now to show that the spectral blocks's properties can be leveraged to design controllers of the type \eqref{eq_main: system_adjacency_controlw'} able to shape the synchronous dynamics of the clusters. More specifically, we will concentrate on three different control tasks. The first deals with the case in which $\mathcal{G}$ does not display spectral blocks in the absence of control (i.e., when $\mathbf{u}_i=0$), and the controllers yield thus the formation of new spectral blocks. The second task is related to the problem of rendering synchronizable clusters of $\mathcal{G}$ (possibly created through the solution of the first task) for a given, desirable, value of the coupling strength $d$. Finally, the third task involves the control of the entire synchronization/desynchronization sequence, namely the order in which the clusters synchronize/desynchronize in class III, as the coupling strength $d$ increases from zero.

For convenience, in what follows the weights ${w'}_{ij}$ are normalized by $d$ (i.e., $w_{ij}=w'_{ij}/d$) so that Eq.~\eqref{eq_main: system_adjacency} is rewritten as
\begin{equation}
    \dot{\mathbf{x}}_i=\mathbf{f}(\mathbf{x}_i)+d \sum_{j=1}^N {a}_{ij}\left(\mathbf{h}(\mathbf{x}_j)-\mathbf{h}(\mathbf{x}_i)\right)+d \sum_{j=1}^N {w}_{ij}\left(\mathbf{h}(\mathbf{x}_j)-\mathbf{h}(\mathbf{x}_i)\right)
    \label{eq_main: system_adjacency_controlw}
\end{equation}
\noindent with $w_{ij}$ being selected such that $w_{ij}+a_{ij}\geq 0$ $\forall i,j=1, \ldots, N$. The adjacency matrix of the controlled network is, therefore, given by $\mathrm{A}'=\mathrm{A}+\mathrm{W}\geq 0$.

The first control task aims at the creation of arbitrary spectral blocks $S_1, S_2, \ldots, S_M$ in a system with dynamics described by Eq.~\eqref{eq_main: system_adjacency_controlw}. Precisely, given $M$ desired clusters of nodes $C_1, C_2, \ldots, C_M$, the problem is to find a control matrix $\mathrm{W}$ such that these clusters are associated with spectral blocks $S_1, S_2, \ldots, S_M$.

To address such a problem, we notice that the condition associating nodes $i$ and $j$ to a spectral block $C_l$ (i.e., $a_{ik}=a_{jk}$ $\forall k \notin C_l$) is in fact similar to the condition guaranteeing that two nodes are symmetric, i.e., $a_{ik}=a_{jk} \forall k=1, \ldots, N$. Therefore, one can follow the approach described in Ref. \cite{gambuzza2020controlling} for inducing symmetries in a graph, and apply it to a fictitious network obtained by neglecting all connections within each cluster. The fictitious network is described by the $N \times N$ adjacency matrix $\mathrm{B}$ with entries $b_{ij}=0$ if $i,j \in C_l$ $\forall l$, and $b_{ij}= a_{ij}$ otherwise.
In order to accomplish the control goal, one can ultimately select the entries of $\mathrm{W}$ such that:
\begin{equation}
    \mathrm{R}_i(\mathrm{B}+\mathrm{W})-(\mathrm{B}+\mathrm{W})\mathrm{R}_i=0 \quad \quad \forall i=1,2,\ldots, M
    \label{eq_main: spectral_block_property}
\end{equation}
\noindent where $\mathrm{R}_i$ is the permutation matrix that maps the nodes of the cluster $C_i=\{i_1, i_2, \ldots, i_{N_i}\}$ into $\{i_2, i_3, \ldots, i_{N_i}, i_1\}$.

In vectorial notation, Eq.~\eqref{eq_main: spectral_block_property} can be rewritten in terms of a system of linear algebraic equations of the type $\mathcal{R} \mathbf{w} = \mathbf{b}$, where $\mathbf{w}$ is a vector of unknown terms (the weights of the control links), and $\mathcal{R}$ and $\mathbf{b}$ are known functions of $\mathrm{B}$ and $\mathrm{R}_1, \mathrm{R}_2, \ldots \mathrm{R}_M$.
As the system of algebraic equations is over-determined one has, in general, infinite solutions, and several optimization conditions can be adopted. In particular, we consider here three cases of: {\it i) } minimizing the norm $\| W\|_2$ (i.e., minimizing the control effort), {\it ii) } maximizing the sparsity of the solution, and {\it iii) } preserving the connectedness of the pristine network (see SM for the exact expression of $\mathcal{R}$ and $\mathbf{b}$, and for more details on the three optimization problems).

\begin{figure} [H]
\centering
 \subfigure[]{\includegraphics[width=.233\textwidth, trim={1.7cm 2cm 2cm 0cm}, clip]{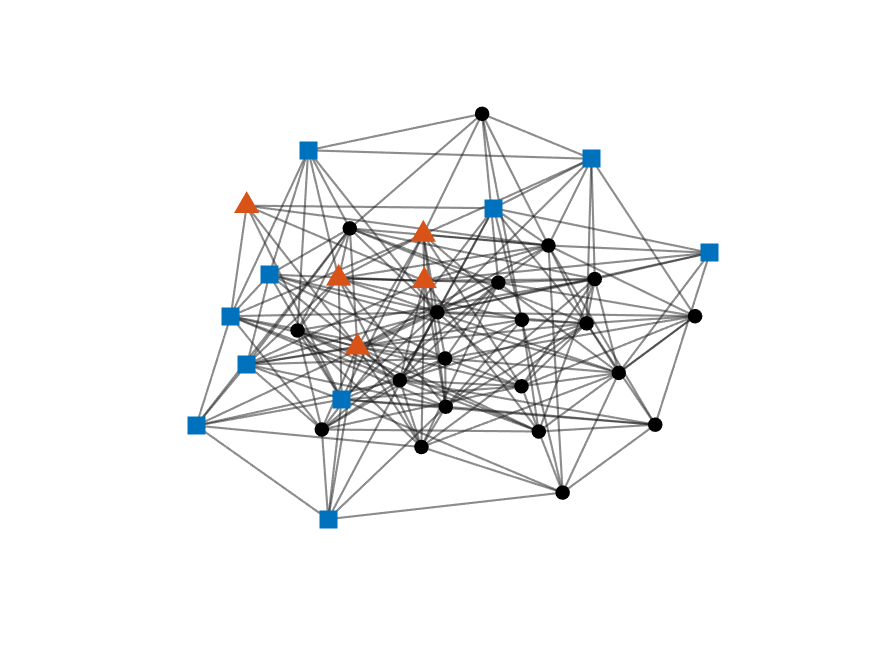} \label{fig: controllo_1a}}
\subfigure[]{\includegraphics[width=.233\textwidth, trim={2cm 2cm 2cm 1.7cm}, clip]{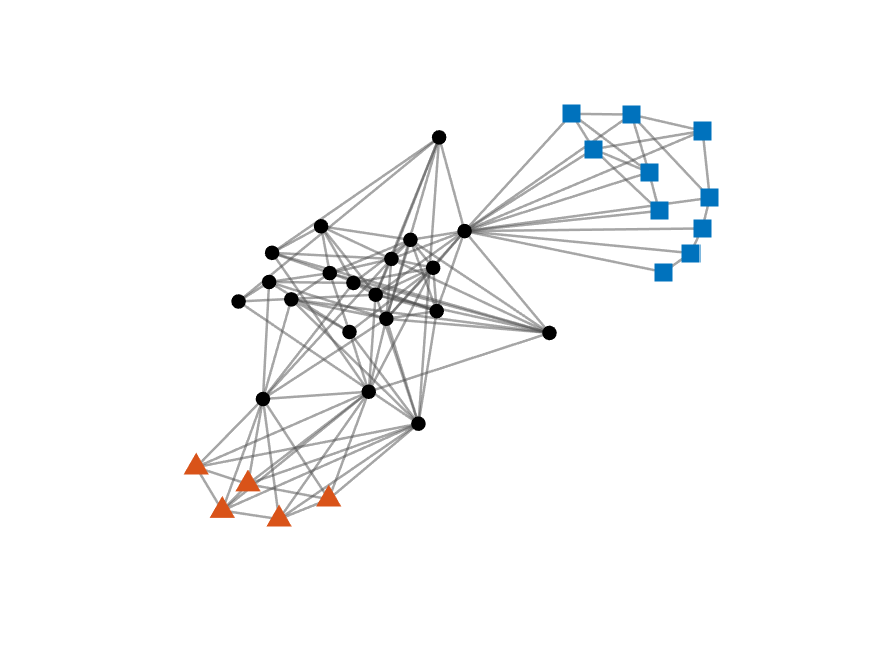} \label{fig: controllo_1b}}
\caption{{\bf Creating spectral blocks.} (a) A graph with $N=35$ nodes and no spectral blocks. The nodes associated with the spectral blocks $S_1$ and $S_2$ to be induced by the control are marked by blue squares and orange triangles, respectively. (b) The controlled network. The control is performed by adding/removing links, and leads to the two desired groups of nodes connected to a bulk with strengths $s_1+w_1=1$ and $s_2+w_2=3$, respectively. \label{fig: controllo1}}
\end{figure}

As an illustrative example, we consider the unweighted graph of Fig.~\ref{fig: controllo_1a}, with $N=35$ nodes. The graph has no spectral blocks, and the task is here to induce two spectral blocks ($S_1$ and $S_2$) formed by the (arbitrarily chosen) nodes marked as blue squares and orange triangles in Fig.~\ref{fig: controllo1}(a), respectively. We proceed with the optimization problem aiming at maximizing the sparsity of the solution (see details in the SM). This either adds or removes links from the pristine network, i.e., $-a_{ij} \leq w_{ij} \leq 1-a_{ij}$, $w_{ij} \in \mathbb{Z}$. The resulting controlled network is depicted in Fig.~\ref{fig: controllo_1b}, where the nodes of $S_1$ and $S_2$ are connected with a bulk of other nodes with strengths $s_1+w_1=1$ and $s_2+w_2=3$, respectively.

The second task consists in considering a graph with $M$ spectral blocks ($S_1, S_2, \ldots, S_M$) associated to the clusters ($C_1, C_2, \ldots, C_M$), and in assuming that $M' \leq M$ of such clusters, namely $C_1, C_2, \ldots, C_{M'}$, cannot synchronize at any value of $d$, as they do not satisfy the eigenvalue ratio condition, i.e. $\frac{\lambda_{N_l}(\mathcal{L}_l)+s_l}{\lambda_{2}(\mathcal{L}_l)+s_l}\geq \frac{\nu_2^*}{\nu_1^*}$ $\forall l=1,\ldots, M'$. The goal is now rendering such clusters synchronizable. For this purpose, control links are added to the network in a way that {\it i)} the spectral block condition on $S_1, S_2, \ldots, S_M$ is preserved, and {\it ii)} the criterion on the eigenvalue ratio becomes fulfilled. In practice, the weights of the control links have to be selected such that $\mathrm{W}=\mathrm{W}^T$, $w_{ik}=w_{jk}$  $\forall i,j \in C_l, \forall k \notin C_l, \forall l=1,\ldots, M$, and:
\begin{equation}
     \frac{\lambda_{N_l}(\mathcal{L}_l)+s_l+w_l}{\lambda_2(\mathcal{L}_l)+s_l+w_l}<\frac{\nu_2^*}{\nu_1^*} \quad \quad \forall l=1,\ldots, M,
\label{eq_main: controlloII_1}
\end{equation}
\noindent where $w_l = \sum _{j\in \mathcal{V} \backslash C_l} w_{ij}$ for $i \in C_l$. Notice that, since
$\lim\limits_{w_l\to\infty} \frac{\lambda_{N_l}(\mathcal{L}_l)+s_l+w_l}{\lambda_2(\mathcal{L}_l)+s_l+w_l}=1<\frac{\nu_2^*}{\nu_1^*}$ $\forall l=1,\ldots,M$, a solution to this problem always exists.
However, also in this case, the problem admits more than one solution, and one has to associate it to an optimization condition. 

In this case, we opted to work with the quotient graph associated to the partition induced by the spectral blocks. Spectral blocks induce a partition of the graph $\pi=\{C_1, C_2, \ldots, C_M, C_{M+1}, \ldots, C_{N_\pi} \}$, where $C_1, C_2, \ldots, C_M$ are the clusters associated with $\mathcal{S}_1, \mathcal{S}_2, \ldots, \mathcal{S}_M$, and $C_{M+1}, \ldots, C_{N_\pi}$ are singletons, each containing one of the remaining nodes. The quotient graph associated to the partition $\pi$, denoted by $\mathcal{G} / \pi$, is the graph with vertices $1, 2, \ldots, {N_\pi}$ and edges connecting nodes $l$ and $m$ with weights $s_{lm}= a_{ij}$, $\forall i \in C_l$, $\forall j \in C_m$. In this way, one can rewrite the inequalities \eqref{eq_main: controlloII_1} in matrix form in terms of the adjacency matrix $\mathrm{S}$ of $\mathcal{G}/ \pi$ and of the $N_\pi \times N_\pi$ matrix $\mathrm{X}$ of elements $x_{lm}=w_{ij}$ ($\forall i \in C_l, \forall j \in C_m, l\neq m$). Working with the quotient graph guarantees that $S_1, S_2, \ldots, S_M$ are spectral blocks also for the controlled network. Finally, by vectorization, one reformulates the optimization problem as a constrained linear inequality where the unknowns are $x_{lm}$ (for $l,m=1, \ldots, M$) and the known terms are the coefficients of $\mathrm{S}$ and the largest and smallest non-zero eigenvalues of $\mathcal{L}_1$ and $\mathcal{L}_2$ (see SM for more details). The solution of the optimization problem provides the weights $w_{ij}$ obtained as $w_{ij}=x_{lm}$ $\forall i \in C_l, \forall j \in C_m$.

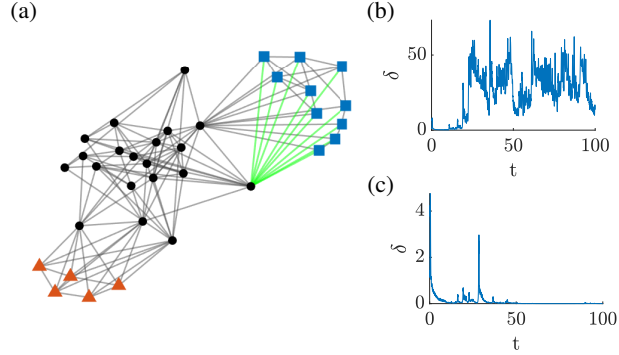
\begin{figure}[H]
\centering
{\begin{tikzpicture}
    \node at (-0.3,-2.2) {\includegraphics[width=0.28\textwidth, trim={2cm 2cm 2cm 1.7cm},clip]{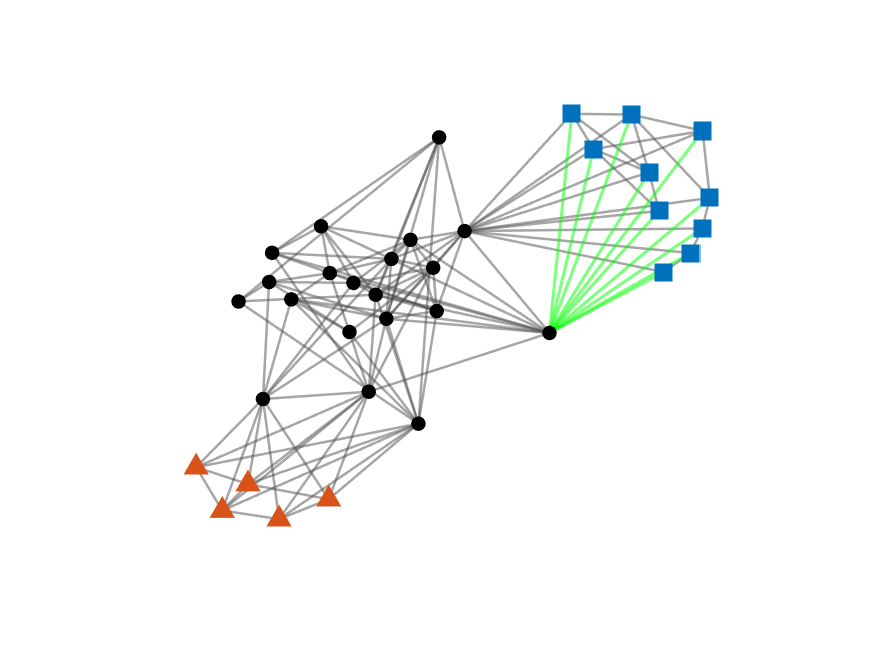}};
    \node at (-2.4,0) {(a)};

    \node at (3.8,-1.1) {\includegraphics[width=0.18\textwidth, trim={-0.5cm 0 0 0},clip]{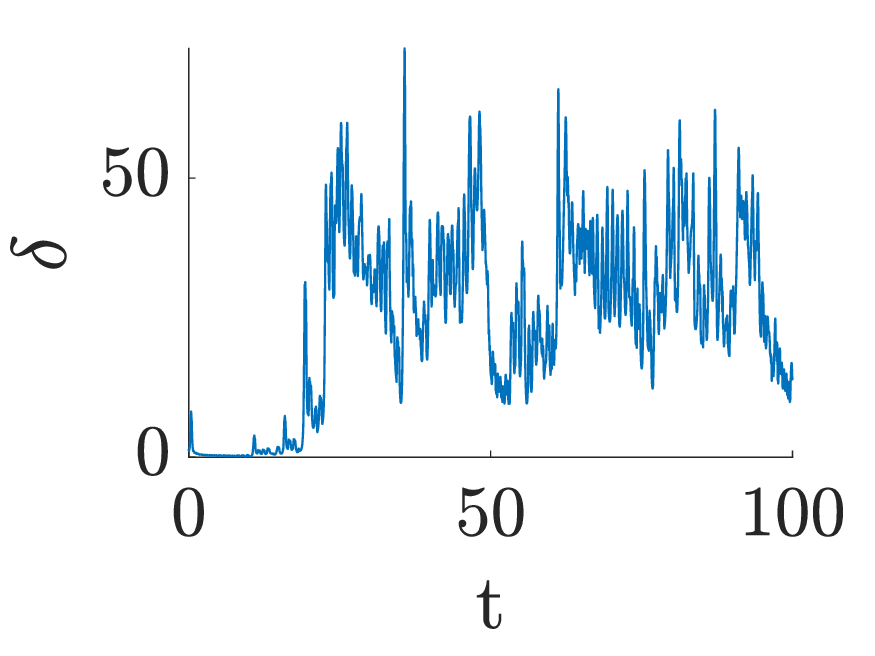}};
    \node at (2.3, 0) {(b)};
    \node at (3.9,-3.4) {\includegraphics[width=0.18\textwidth, trim={-0.5cm 0 0 0},clip]{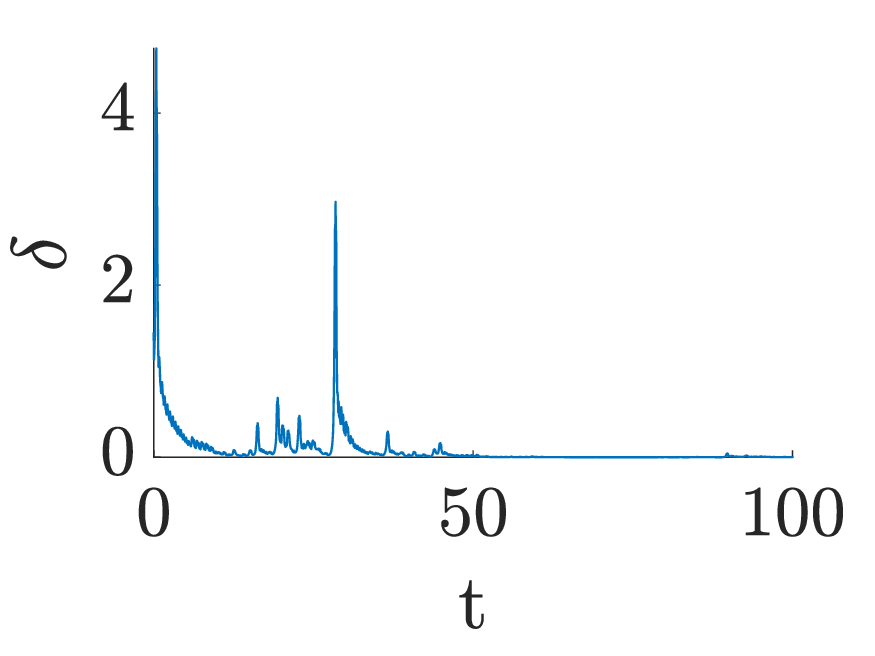}};
    \node at (2.3, -2.3) {(c)};
  \end{tikzpicture}}
\caption{{\bf Taming cluster synchronizability.} (a) Controlled network with two synchronizable spectral blocks $S_1$ and $S_2$. The nodes where the spectral blocks are localized are now connected with the bulk with strengths $s_1+w_1=2$ and $s_2+w_2=3$, respectively. (b) Time evolution of the synchronization error $\delta(t)$ within the cluster $C_1$, for $d=2$ in the absence of control (i.e., using the network of Fig.~\ref{fig: controllo_1b}). (c) Time evolution of $\delta(t)$ within the cluster $C_1$, for $d=2$ when control is applied (i.e., using the network of panel (a)). \label{fig: controllo2}}
\end{figure}

As an example, we consider the graph of Fig.~\ref{fig: controllo_1b} i.e., the result of the first control task, displaying two spectral blocks ($S_1, S_2$) with associated clusters $C_1, C_2$. Without lack of generality, we consider the node dynamics regulated by the Lorenz system \cite{lorenz1963deterministic}. Therefore Eqs.~\eqref{eq_main: system_adjacency_controlw} read:
\begin{equation}
    \begin{cases}
        \dot{x}_{i,1}= \sigma \left(x_{i,2}-x_{i,1} \right)+ d \sum\limits_{j=1}^N \left({a}_{ij}+w_{ij} \right) \left({x}_{j,2}-{x}_{i,2} \right),\\
        \dot{x}_{i,2}= x_{i,1}\left(\rho - x_{i,3} \right)-x_{i,2},\\
        \dot{x}_{i,3}= x_{i,1} x_{i,2}- \beta x_{i,3},
    \end{cases}
\end{equation}
\noindent where the parameters are $\sigma=10$, $\rho=28$, and $\beta=2$, so as the uncoupled dynamics is chaotic. The system has a type III master stability function with $\nu_1^*=4.173$ and $\nu_2^*=22.535$ \cite{huang2009generic}. The smallest and the largest non-zero eigenvalues of $\mathcal{L}_1$ are $\lambda_2(\mathcal{L}_1)=0.21$ and $\lambda_{10}(\mathcal{L}_1)=5.93$, whereas those of $\mathcal{L}_2$ are $\lambda_2(\mathcal{L}_2)=1.38$ and $\lambda_5(\mathcal{L}_2)=4.62$ (see SM for all details on $\mathcal{L}_1$ and $\mathcal{L}_2$). Furthermore, the clusters $C_1$ and $C_2$ are connected to the rest of the graph with strengths $s_1=1$ and $s_2=3$, respectively. Since $\frac{\lambda_{10}(\mathcal{L}_1)+s_1}{\lambda_{2}(\mathcal{L}_1)+s_1}>\frac{\nu_2^*}{\nu_1^*}$, the nodes of $C_1$ cannot synchronize at any value of $d$. For instance, a large synchronization error $\delta=\frac{1}{N_1} \left( \sum_{i \in C_1} ||\mathbf{x}_i-\bar{\mathbf{x}}_1||^2 \right)^{\frac{1}{2}}$ (with $N_1=|C_1|$ and $\bar{\mathbf{x}}_1=\frac{1}{N_1}\sum_{j \in C_1} \mathbf{x}_j$) is obtained for $d=2$, as shown in Fig.~\ref{fig: controllo2}(b). Also in this case, we adopt the optimization problem that maximizes the sparsity of the solution. Fig.~\ref{fig: controllo2}(a) shows the controlled network, in which $C_1$ is connected to the bulk with a strength $s_1+w_1=2$. Since $\frac{\lambda_{10}(\mathcal{L}_1)+s_1+w_1}{\lambda_{2}(\mathcal{L}_1)+s_1+w_1}<\frac{\nu_2^*}{\nu_1^*}$, $C_1$ now satisfies the eigenvalue ratio condition and can therefore synchronize for $1.89=\frac{\nu_1^*}{\lambda_2(\mathcal{L}_1)+s_1+w_1}<d<\frac{\nu_2^*}{\lambda_{10}(\mathcal{L}_1)+s_1+w_1}=2.84$. Consequently, $C_1$ reaches synchronization for $d=2$, as confirmed from the time evolution of the error $\delta(t)$ shown in Fig.~\ref{fig: controllo2}(c).

The third control task considers a graph equipped with $M$ spectral blocks ($\mathcal{S}_1, \mathcal{S}_2, \ldots, \mathcal{S}_M$) forming the clusters $C_1, C_2, \ldots, C_M$. Let $s_1, s_2, \ldots, s_M$ and $\mathcal{L}_1, \mathcal{L}_2, \ldots, \mathcal{L}_M$ be, respectively, the strengths through which the clusters are connected with the rest of the graph and the Laplacian matrices of the subgraphs associated with the clusters. One can define the synchronization sequence $\mathcal{I}$ as the set that contains the cluster indices $j$, ordered in decreasing order with respect to the sum of the largest non-zero eigenvalue of $\mathcal{L}_{j}$ and the corresponding $s_{j}$. In other words, $\mathcal{I}=\{i_1, i_2, \ldots, i_M \}$ if $\lambda_2(\mathcal{L}_{i_1})+s_{i_1} \geq \lambda_2(\mathcal{L}_{i_2})+s_{i_2} \geq \ldots \geq \lambda_2(\mathcal{L}_{i_M})+s_{i_M}$. Similarly, one may define the desynchronization sequence $\mathcal{D}$ as  $\mathcal{D}=\{i_1, i_2, \ldots, i_M \}$ if $\lambda_{N_{i_1}}(\mathcal{L}_{i_1})+s_{i_1} \geq \lambda_{N_{i_2}}(\mathcal{L}_{i_2})+s_{i_2} \geq \ldots \geq \lambda_{N_{i_M}}(\mathcal{L}_{i_M})+s_{i_M}$, with $N_i=|C_i|$. Clusters, indeed, will enter (exit) one after the other the stability region following the order specified in the sequence $\mathcal{I}$ ($\mathcal{D}$).

The goal of the control is to induce synchronization and desynchronization sequences chosen {\it ad libitum}, say $\mathcal{I}'=\mathcal{D}'=\{1, 2, \ldots, M\}$. To that purpose, the weights of the control links have to be selected such that: {\it i)} the clusters $C_1, C_2, \ldots, C_M$ remain associated with the spectral blocks $S_1, S_2, \ldots, S_M$, {\it ii)} they satisfy the eigenvalue ratio condition, {\it iii)} they synchronize and desynchronize according to the desired sequences $\mathcal{I}'$ and $\mathcal{D}'$. In practice, the entries of $\mathrm{W}$ have to be selected such that
Eq.~\eqref{eq_main: controlloII_1} holds, $\mathrm{W}=\mathrm{W}^T$, $w_{ik}=w_{jk}$  $\forall i,j \in C_l, \forall k \notin C_l, \forall l=1,\ldots, M$, and:
\begin{equation}
\begin{dcases}
        \frac{\nu_1^*}{\lambda_2(\mathcal{L}_l)+s_l+w_l} < \frac{\nu_1^*}{\lambda_2(\mathcal{L}_{l+1})+s_{l+1}+w_{l+1}} \\
        \frac{\nu_2^*}{\lambda_{N_l}(\mathcal{L}_l)+s_l+w_l} < \frac{\nu_2^*}{\lambda_{N_{l+1}}(\mathcal{L}_{l+1})+s_{l+1}+w_{l+1}}
        \label{eq_main: controlloIII}
\end{dcases}
\end{equation}
\noindent $\forall l=1,\ldots, M-1$.

Also in this case the weights $w_{ij}$ associated to the controllers are found solving an optimization problem. Namely, one starts from considering the quotient graph $\mathcal{G} / \pi$ and its adjacency matrix $\mathrm{S}$, and rewrites the inequalities~\eqref{eq_main: controlloIII} in matrix form through $\mathrm{S}$ and $\mathrm{X}$. Then, via vectorization, one ends up with a constrained linear inequality where the unknown terms are the entries of $\mathrm{X}$, whereas the known terms depend on $\mathrm{S}$ and the smallest and largest non-zero eigenvalues of $\mathcal{L}_1$ and $\mathcal{L}_2$ (see SM for the details). Eventually, a set of weights $w_{ij}$ satisfying such a constrained linear inequality can be found by solving optimization problems where the $L_2$ norm of the solution is minimized, or the sparsity of the matrix is maximized, or the connectedness of the structure is preserved.

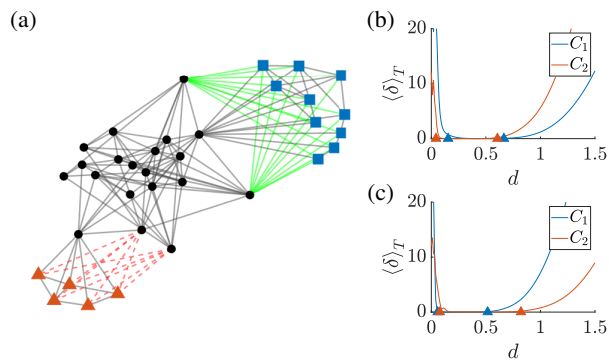
\begin{figure}[h]
\centering

{\begin{tikzpicture}
    \node at (-0.3,-2.2) {\includegraphics[width=0.28\textwidth, trim={2cm 2cm 2cm 1.7cm},clip]{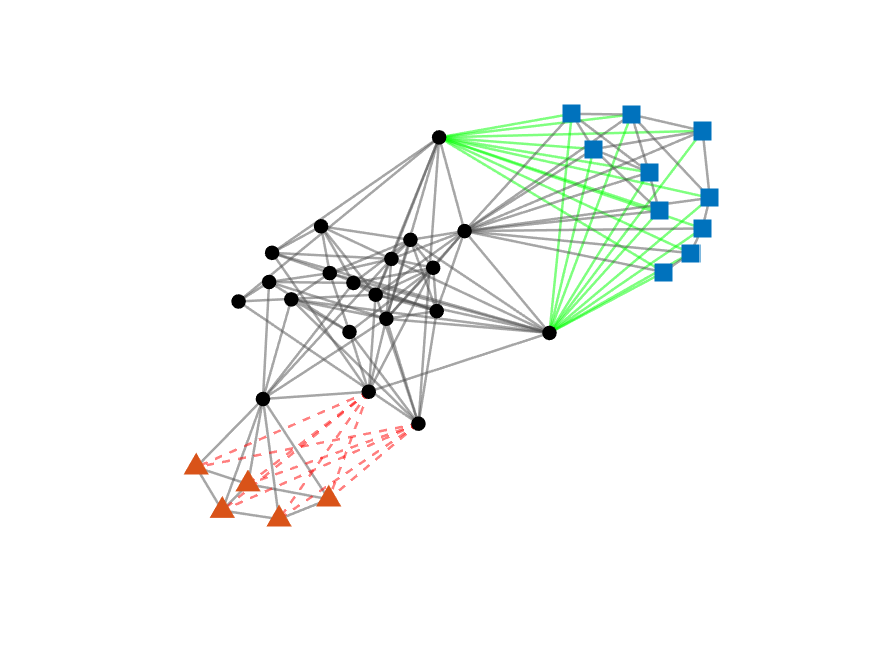}};
    \node at (-2.4,0) {(a)};

    \node at (3.8,-1.1) {\includegraphics[width=0.18\textwidth, trim={-0.5cm 0 0 0},clip]{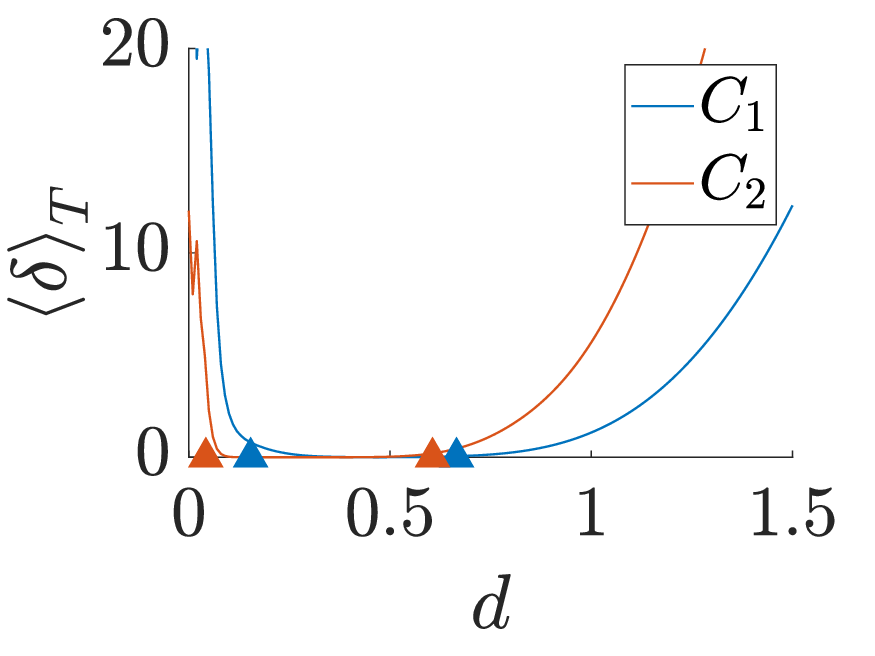}};
    \node at (2.3, 0) {(b)};
    \node at (3.8,-3.4) {\includegraphics[width=0.18\textwidth, trim={-0.5cm 0 0 0},clip]{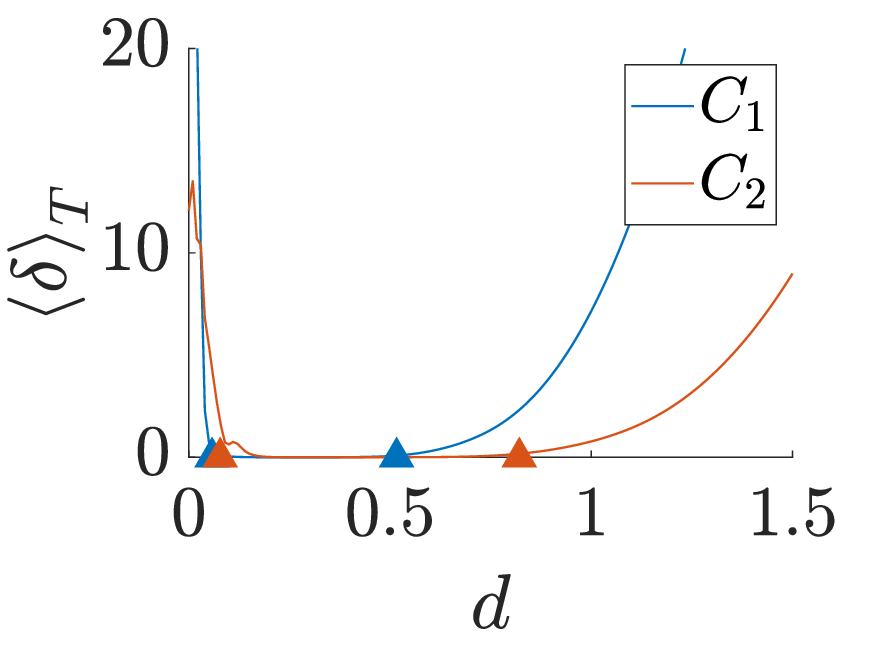}};
    \node at (2.3, -2.3) {(c)};
  \end{tikzpicture}}
\caption{{\bf Shaping the synchronization/desynchronization sequence.} (a) Controlled network.  The clusters $C_1$ and $C_2$ are connected with the bulk nodes with strengths $s_1+w_1=3$ and $s_2+w_2=1$. (b, c) $\langle \delta_h \rangle$ (see text for definition) vs. $d$ for the uncontrolled [Fig.~\ref{fig: controllo_1b}] and controlled [panel (a)] networks, relative to the clusters $C_1$ (blue curve) and $C_2$ (red curve). The triangles (blue for $C_1$ and orange for $C_2$) in the horizontal axes mark the critical values for synchronization as predicted by the MSF. It is seen that the effect of control is to switch the sequence from $\mathcal{I}=\mathcal{D}=\{2,1\}$ to $\mathcal{I'}=\mathcal{D'}=\{1,2\}$.
\label{fig: controllo3}}
\end{figure}

In order to show the effectiveness of the proposed control, let us focus on the network of Fig.~\ref{fig: controllo_1b} endowed with the two clusters $C_1$ and $C_2$. To show how general the applicability of our method is, this time we take for each node the dynamics of the R\"ossler oscillator \cite{rossler1976equation}. The network evolution is therefore governed by the following equations:
\begin{equation}
\begin{cases}
\dot{x}_{i,1}=-{x}_{i,2}-{x}_{i,3}+d \sum\limits_{j=1}^N \left({a}_{ij}+w_{ij} \right) \left({x}_{j,1}-{x}_{i,1} \right),\\
\dot{x}_{i,2}={x}_{i,1}+a{x}_{i,2},\\
\dot{x}_{i,3}=b+{x}_{i,3}({x}_{i,1}-c),\\
\end{cases}
\label{eq_main:RosslerCasoMSFTipoIII}
\end{equation}

\noindent with $a=0.2$, $b=0.2$ and $c=7$, such that the uncoupled dynamics is chaotic. System~\eqref{eq_main:RosslerCasoMSFTipoIII} has a type III MSF with $\nu_1^*=0.186$ and $\nu_2^*=4.614$ \cite{huang2009generic}. Here, the connections from $C_1$ and $C_2$ to the rest of the graph are such that $s_1=1$ and $s_2=3$. Under these conditions, the smallest and the largest non-zero eigenvalues of $\mathcal{L}_1$ are $\lambda_2(\mathcal{L}_1)=0.21$ and $\lambda_{10}(\mathcal{L}_1)=5.93$, and those of $\mathcal{L}_2$ are $\lambda_2(\mathcal{L}_2)=1.38$ and $\lambda_5(\mathcal{L}_2)=4.62$. One then has that $\lambda_2(\mathcal{L}_2)+s_2 > \lambda_2(\mathcal{L}_1)+s_1$ and $\lambda_5(\mathcal{L}_2)+s_2>\lambda_{10}(\mathcal{L}_1)+s_1$, and therefore, in the absence of control, the synchronization and desynchronization sequence are $\mathcal{I}=\mathcal{D}= \{2, 1\}$. This is confirmed by the numerical simulations illustrated in Fig. \ref{fig: controllo3}(b), which shows the average value of the synchronization error over a time window $T=10$ within the interval $[4T, 5T]$, i.e., $\langle \delta_h \rangle_T=
\langle \frac{1}{N_h} \left( \sum_{i \in C_h} ||\mathbf{x}_i^2-\mathbf{\bar{x}_h}||^2\right)^{\frac{1}{2}} \rangle_T$, for the two clusters $C_1$ and $C_2$, i.e., $h=\{1,2\}$  vs. the coupling strength $d$.
The critical values predicted by the approach described in Ref. \cite{bayani2023transition} are also reported in Fig.~\ref{fig: controllo3}(b) as blue (orange) triangles for $C_1$ ($C_2$).

The performed control aims at changing the synchronization/desynchronization sequence into $\mathcal{I}'=\mathcal{D}'=\{1,2\}$, and the corresponding $\mathrm{W}$ can be found by fulfilling the conditions in Eq.~\eqref{eq_main: controlloIII}. This yields the controlled network shown in Fig.~\ref{fig: controllo3}(a) with the clusters $C_1$ and $C_2$ being connected to the bulk with strengths $s_1+w_1=1$ and $s_2+w_2=3$, respectively. Consequently, we have that $\lambda_2(\mathcal{L}_1)+s_1+w_1>\lambda_2(\mathcal{L}_2)+s_2+w_2$ and $\lambda_5(\mathcal{L}_2)+s_2+w_2<\lambda_{10}(\mathcal{L}_1)+s_1+w_1$, resulting in $\mathcal{I}'=\mathcal{D}'=\{1,2\}$.  $\langle \delta_h \rangle$  vs. $d$ for $C_1$ and $C_2$ ($h=\{1,2\}$), is shown in Fig.~\ref{fig: controllo3}(c), and confirms that the controlled network displays the imprinted synchronization/desynchronization sequence. In this case, we searched for a solution that only adds or remove links, without changing the weights of existing links and ensuring that both clusters remain connected to the bulk (see SM for the implementation details).

In conclusion, we have shown that spectral blocks can be efficiently used to fully shape the dynamics of synchronous clusters within a generic network of dynamical units. This implies that, through the judicious deployment of properly designed controllers, one is able to exercise the authority to manipulate cluster synchronization at will and with precision, and to gain unprecedented mastery over the synchronization landscape of networks.

Our results are of value and relevance in several areas of physics. The ability to manage the entry and exit of clusters in and out of synchronization, in fact, represents a veritable linchpin for orchestrating the parallel behavior and functioning of distributed systems. This transformative capacity not only allows the efficiency and adaptability of such systems to be amplified, but also opens avenues for exploring new applications in biological and technological networks.  By elucidating how cluster synchronization control can be made, our work underscores the potential for facilitating the optimization of complex network's dynamics for real-world applications.

\clearpage

\section*{Supplemental Materials}

This supplementary material provides a detailed description of all methods for spectral block control used in the Main Text, as well as additional numerical examples.

\section*{Mathematical preliminaries}

\textit{Notation.} Here and in the Main Text, we use the following notation: $\mathrm{I}_N$ indicates the $N \times N$ identity matrix, $\mathrm{0}_{N,M}$ is a $N \times M$ matrix with all entries equal to 0, $\mathbf{1}_N$ denotes a $N-$dimensional vector with all entries equal to 1. Given a $N \times M$ matrix $\mathrm{A}$, we denote with ${\rm vec}(\mathrm{A})$ the vectorization of $\mathrm{A}$ obtained by stacking the columns of $\mathrm{A}$ as follows: $${\rm vec}(\mathrm{A})=[a_{1,1}, a_{2,1}, \ldots, a_{N,1}, a_{1,2}, a_{2,2}, \ldots, a_{N,2}, \ldots, a_{1,M}, a_{2,M}, \ldots, a_{N,M}]^T$$
\noindent where $a_{ij}$, with $i=1,\ldots,N$, $j=1,\ldots,N$, are the coefficients of $\mathrm{A}$ \cite{abadir2005matrix}.
Finally, given a $N \times N$ matrix $\mathrm{A}$, ${\rm diag}(\mathrm{A})$ is a $N\times N$ matrix having in the diagonal the elements of the diagonal of $\mathrm{A}$ and zero elsewhere.

\textit{Graphs.} A graph is a mathematical structure described by the pair $\mathcal{G} = (\mathcal{V}, \mathcal{E})$, where $\mathcal{V}(\mathcal{G})=\{1, 2, \dots, N\}$ is the set of the vertices/nodes, and $\mathcal{E}(\mathcal{G}) \subseteq \mathcal{V} \times \mathcal{V}$ the set of edges, that are ordered pairs of vertices. A graph is undirected if, for any $(i,j) \in \mathcal{E}$, there exists $(j,i) \in \mathcal{E}$. A graph can be represented by the adjacency matrix $\mathrm{A}$ and the Laplacian matrix $\mathcal{L}$. The adjacency matrix $\mathrm{A}$ is a $N \times N$ matrix with $a_{ij}>0$ if there is an edge between $i$ and $j$, and $a_{ij}=0$ otherwise. More specifically, $a_{ij}$ represents the weight of the edge between the nodes $i$ and $j$, $\forall (i,j) \in \mathcal{E}$. The Laplacian matrix $\mathcal{L}$ is a $N \times N$ zero-row sum matrix with entries $\mathcal{L}_{ij}=-a_{ij}$ if $i \neq j$ and $\mathcal{L}_{ii}=\sum_{i=1}^N a_{ij}$. In particular, the Laplacian matrix is positive semidefinite.  When the graph is undirected, we have $a_{ij}=a_{ji}$ and $\mathcal{L}_{ij}=\mathcal{L}_{ji}$ $\forall i,j=1, \ldots, N$. When the graph is undirected and connected, the eigenvalues of $\mathcal{L}$ are all non-negative, and only one is equal to zero, such that they can be ordered as $0=\lambda_1(\mathcal{L})<\lambda_2(\mathcal{L}) \leq \lambda_3(\mathcal{L}) \ldots \leq \lambda_N(\mathcal{L})$. A graph is said to be simple if it does not contain self-loops, i.e. $a_{ii}=0$ $\forall i =1, \ldots, N$. In this work, we consider graphs that are undirected, connected and simple.

\textit{Spectral blocks.} A spectral block $\mathcal{S}$ localized at nodes $\{i_1, i_2, \ldots, i_{N'}\}$ is defined as a subset of $(N'-1)$ eigenvectors of the Laplacian matrix ${\cal L}$ having the following properties \cite{bayani2023transition}:  {\it i) } all $\mathbf{v} \in \mathcal{S}$ are such that $\nu_i=0$ $\forall i \notin {i_1, i_2, \ldots, i_N'}$;  {\it ii) } all $\mathbf{v} \notin \mathcal{S}$ are such that $\nu_i=\nu_j$ $\forall i,j \in {i_1, i_2, \ldots, i_N'}$. As shown in \cite{bayani2023transition}, a group of nodes $C$ is associated with a spectral block $\mathcal{S}$ if and only if they are equally connected (i.e., with the same weight) to each other node that is not in $C$, i.e. $a_{ik}=a_{jk}$, $\forall i,j \in C$ and $\forall k \notin C$.

When a graph is equipped with the spectral blocks $\mathcal{S}_1, \mathcal{S}_2, \ldots, \mathcal{S}_M$, a partition $\pi=\{C_1, C_2, \ldots, C_{M}, C_{M+1}, \ldots C_{N_\pi} \}$ can be considered, where $C_1, C_2, \ldots, C_M$ are the clusters associated with $\mathcal{S}_1, \mathcal{S}_2, \ldots, \mathcal{S}_M$, and $C_{M+1}, \ldots, C_{N_\pi}$ are singletons, each containing one of the remaining nodes.
This partition can be associated with a quotient graph, denoted by $\mathcal{G} / \pi$, that has vertices $1, 2, \ldots, {N_\pi}$ and edges with weights $s_{lm}= a_{ij}$ connecting nodes $l$ and $m$ $\forall i \in C_l$, $\forall j \in C_m$. The quotient graph can be represented by its adjacency matrix $S=\{s_{lm}\}$ with $l,m=1,\ldots, N_\pi$.

The following lemma will be useful in the analysis of the control problems discussed below:
\begin{lemma}{Farka's lemma \cite{matouvsek2007understanding}.}
    Let $\mathrm{A} \in \mathbb{R}^{m \times n}$ and $\mathbf{b} \in \mathbb{R}^m$, then only one of the following statements is true:
\begin{itemize}
    \item There exists $\mathbf{x} \geq 0$ such that $\mathrm{A} \mathbf{x}\leq \mathbf{b}$.
    \item There exists $\mathbf{y} \geq 0$ such that $\mathrm{A}^T \mathbf{y} \geq 0$ and $\mathbf{b}^T \mathbf{y}>0$.
    \end{itemize}
    \label{lemma: Farkas}
\end{lemma}

\section*{The Master Stability Function}

Let us consider a network of $N$ coupled oscillators described by the following dynamics:
\begin{equation}
    \dot{\mathbf{x}}_i=\mathbf{f}(\mathbf{x}_i)+d \sum_{j=1}^N a_{ij}\left(\mathbf{h}(\mathbf{x}_j)-\mathbf{h}(\mathbf{x}_i)\right)
    \label{eq: system}
\end{equation}
where $i=1, \ldots, N$, $\mathbf{f}$ is the uncoupled dynamics, $d$ is the coupling strength, $a_{ij}$ are the entries of the adjacency matrix describing the interaction graph $\mathcal{G}$, which is assumed to be undirected and weighted, and $\mathbf{h}$ is the inner coupling function.

Eq.~\eqref{eq: system} is such that the synchronization manifold, that is defined by $\mathbf{x}_1=\mathbf{x}_2= \ldots=\mathbf{x}_N=\mathbf{x}_s$, always exists and has dynamics described by the following equation:
\begin{equation}
    \dot{\mathbf{x}}_s=\mathbf{f}(\mathbf{x}_s)
\end{equation}
To study the stability of this solution, we follow the approach introduced in \cite{pecora1998master}, which starts by considering a small perturbation $\delta \mathbf{x}_i = \mathbf{x}_i-\mathbf{x}_s$ around the synchronization manifold, and by linearizing the system dynamics around the synchronization manifold:

\begin{equation}
    \dot{\delta {\mathbf{x}}}= \left[\mathrm{I} \otimes D\mathbf{f}|_{\mathbf{x}_s} - d \mathcal{L} \otimes D\mathbf{h}|_{\mathbf{x}_s} \right]\delta {\mathbf{x}}
    \label{eq: variational_equation}
\end{equation}

\noindent where $\delta {\mathbf{x}} = [\delta \mathbf{x}_1^T, \delta \mathbf{x}_2,^T \ldots, \delta \mathbf{x}_N^T]^T$. Here, $\mathrm{D} \mathbf{f}|_{\mathbf{x}_s}$ and $\mathrm{D} \mathbf{h}|_{\mathbf{x}_s}$ are the Jacobian matrix of $\mathbf{f}$ and $\mathbf{h}$ computed around the synchronous manifold $\mathbf{x}_s$, respectively. Eq.~\eqref{eq: variational_equation} is then block-diagonalized resulting in a new set of equations where each block has the form $\dot{\mathbf{\xi}}_i=\left[\mathrm{D}  \mathbf{f}|_{\mathbf{x}_s} - d \lambda_i(\mathcal{L}) \mathrm{D}\mathbf{h}|_{\mathbf{x}_s} \right] \mathbf{\xi}_i$. Since the blocks only differ for the eigenvalue appearing in it, by introducing the parameter $\nu=d\lambda_i(\mathcal{L})$, a single Master Stability Equation (MSE), namely $\dot{\mathbf{\zeta}}=\left[\mathrm{D}  \mathbf{f}|_{\mathbf{x}_s} - \nu \mathrm{D}\mathbf{h}|_{\mathbf{x}_s} \right]\mathbf{\zeta}$, can be considered. This is an important step, as it allows to separate the role of the unit dynamics (namely $\mathbf{f}$ and $\mathbf{h}$) from that of the structure of interactions (the eigenvalues of the Laplacian matrix) in the variational equation.  From the MSE, the maximum Lyapunov exponent $\lambda_{\rm max}$ is calculated as a function of $\nu$, thus obtaining the Master Stability Function (MSF), i.e., $\lambda_{\rm max}=\lambda_{\rm max}(\nu)$. The condition on stability of synchronization is then expressed as $\lambda_{\rm max}(\nu)=\lambda_{\rm max}(d\lambda_i(\mathcal{L}))<0 $ $\forall i=2, \ldots, N$. As a result, the MSF provides a necessary condition for the stability of the synchronization manifold that is effective and easy to check, unveiling how network topology affects the property of synchronization stability.

\begin{figure}[H]
\centering
\includegraphics[width=0.4\textwidth]{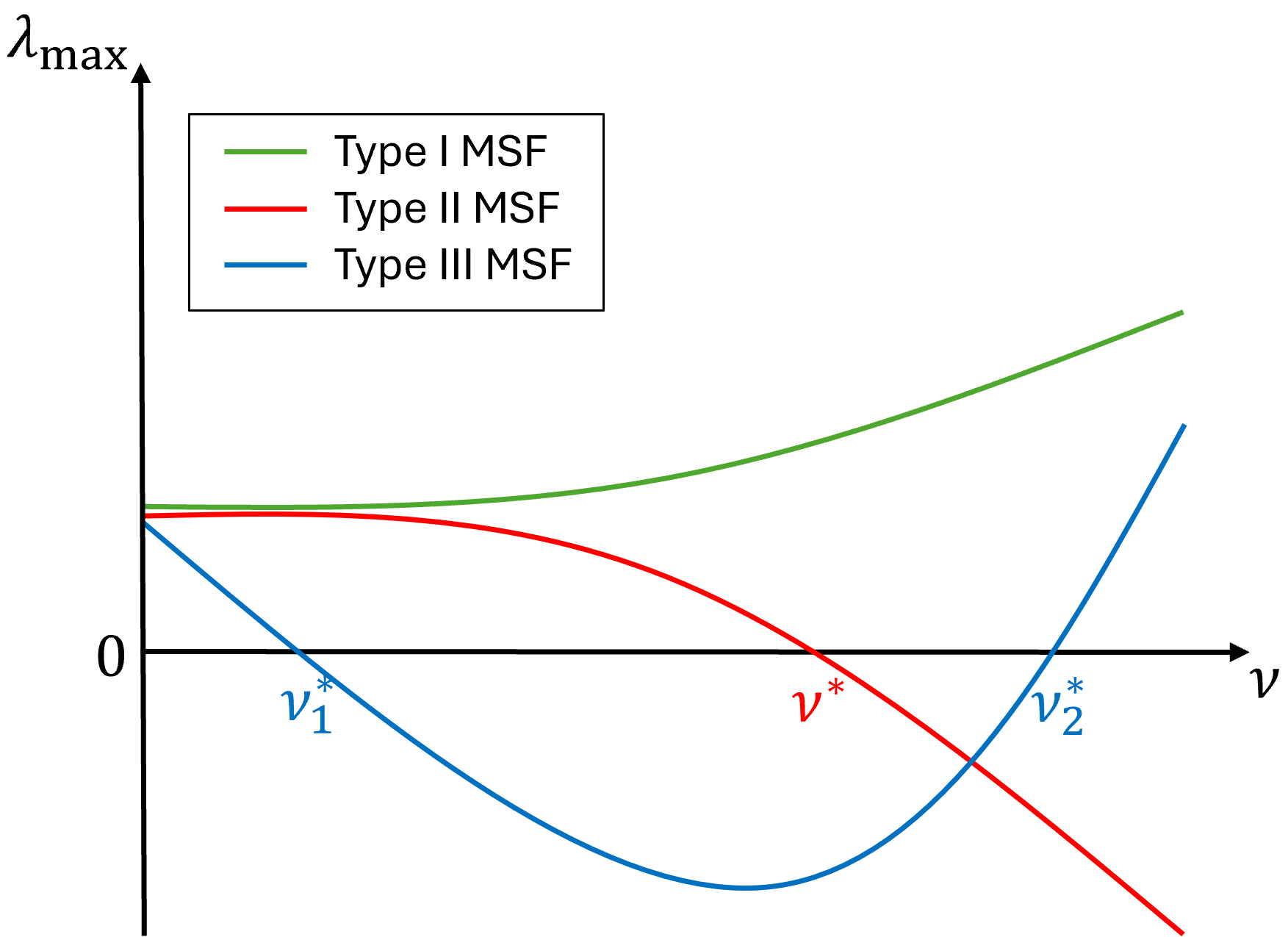}
\caption{Possible classes of MSF. \label{fig:MSF}}
\end{figure}

In \cite{boccaletti2006synchronization}, three classes of MFSs have been identified for chaotic systems: type I MSF, type II MSF, and type III MSF. In systems with type I MFS, $\lambda_{\rm max}(\nu)>0$ for any value of $\nu$, making the synchronization manifold unstable $\forall d$. In systems with type II MSF, $\lambda_{\rm max}(\nu)$ turns from positive to negative values at the critical value $\nu^*$, yielding a scenario where a transition from instability to stability of the synchronization manifold can be observed when the coupling strength is increased from zero. In this case, synchronization stability can be achieved if $d\lambda_2(\mathcal{L})>\nu^*$. For systems with type III MSF, $\lambda_{\rm max}(\nu)<0$ for $\nu \in [\nu_1^*, \nu_2^*]$, where $\nu_1^*$ and $\nu_2^*$ are two threshold values. This yields the possibility of observing two transitions when the coupling strength is varied from zero: from instability to stability and from stability to instability. Also for systems with type III MSF, synchronization stability depends on the spectrum of $\mathcal{L}$; more specifically, it requires that $d \lambda_i \in [\nu_1^*, \nu_2^*]$ $\forall i=2, \ldots, N$. This condition can be achieved only if $\frac{\lambda_N(\mathcal{L})}{\lambda_2(\mathcal{L})} < \frac{\nu_2^*}{\nu_1^*}$. Fig.~\ref{fig:MSF} summarizes the three classes of MSF that can be observed.

The approach based on the MSF can be expanded to encompass the study of cluster synchronization \cite{bayani2023transition}. Here we consider the case in which the clusters are induced by the presence of spectral blocks. Specifically, given a cluster $C$ formed by $M$ nodes, taking into account the defining structural property of a spectral block, for each node $l \in C$, Eq.~\eqref{eq: system} can be rewritten as follows:
\begin{equation}
    \dot{\mathbf{x}}_l=\mathbf{f}(\mathbf{x}_l)-d \sum_{m \in C} l_{lm} \mathbf{h}(\mathbf{x}_m)-d \sum_{m \notin C} l_{lm} \mathbf{h}(\mathbf{x}_m)
    \label{eq: system2}
\end{equation}
\noindent where the coupling term is split into two sums, one including extra-cluster interactions and one including intra-cluster interactions.
The dynamics of the cluster synchronous solution, defined by $\mathbf{x}_l=\mathbf{x}_m= \ldots = \mathbf{x}_{C}$ $\forall l,m \in C$, is given by the following equation:
\begin{equation}
    \dot{\mathbf{x}}_C=\mathbf{f}(\mathbf{x}_C)+d \sum_{m \notin C} a_{lm} \left( \mathbf{h}(\mathbf{x}_m)-\mathbf{h}(\mathbf{x}_C)\right)
    \label{eq: cluster_synchronous_dynamics}
\end{equation}

By considering the perturbation around the cluster synchronous state $\delta \mathbf{x}_{l,C} = \mathbf{x}_l - \mathbf{x}_C$ and by performing linearization of Eq.~\eqref{eq: system2}, we obtain:
\begin{equation}
    \dot{\delta {\mathbf{x}_C}}=\left[\mathrm{I} \otimes D \mathbf{f} |_{\mathbf{x}_C}-d \mathcal{L} \otimes D\mathbf{h}|_{\mathbf{x}_C} \right] \delta {\mathbf{x}_C}
    \label{eq: cluster_synchronous_error}
\end{equation}

\noindent with $\delta \mathbf{x}_C=[\delta \mathbf{x}_{1,C}^T, \delta \mathbf{x}_{2,C}^T, \ldots, \delta \mathbf{x}_{N,C}^T]^T$. Eq.~\eqref{eq: cluster_synchronous_error} is very similar to Eq.~\eqref{eq: variational_equation}, with the difference that, in this case, the Jacobians of $\mathbf{f}$ and $\mathbf{g}$ are evaluated around the cluster synchronous solution $\mathbf{x}_C$, whose dynamics is defined by Eq.~\eqref{eq: cluster_synchronous_dynamics}. In this scenario, the trajectories of the cluster synchronous state are affected by the last term of Eq.~\eqref{eq: cluster_synchronous_dynamics}, therefore they also depend on the dynamics of the rest of the network. Here, we assume that this term has a negligible effect on the trajectories of the cluster synchronous state, or more precisely on the maximum transverse Lyapunov exponent that results from the use of such trajectories in Eq.~\eqref{eq: cluster_synchronous_error}. Under this assumption, we can replace the trajectories followed by the clustered synchronous nodes with the ones of global synchronization in Eq.~\eqref{eq: cluster_synchronous_error}. This allows assessing the stability of each cluster's synchronous state by studying the MSF associated with global synchronization.




\section*{Controlling the network}
\label{control}
In this section, we consider the problem of controlling the networked dynamical system \eqref{eq: system}. To this aim, we introduce a control input $\mathbf{u}_i$ in the equations governing the system dynamics:
\begin{equation}
    \dot{\mathbf{x}}_i=\mathbf{f}(\mathbf{x}_i)+d \sum_{j=1}^N a_{ij}\left( \mathbf{h}(\mathbf{x}_i) -\mathbf{h}(\mathbf{x}_j)\right)
    +\mathbf{u}_i
    \label{eq: dynamics}
\end{equation}
where the control input $\mathbf{u}_i$ is assumed to take the following form:
\begin{equation}
    \mathbf{u}_i = \sum_{j=1}^N w'_{ij}\left( \mathbf{h}(\mathbf{x}_i) -\mathbf{h}(\mathbf{x}_j)\right)
\end{equation}
\noindent We note that the control acts introducing further links in the structure. These links operate in the same fashion as those of the pristine network, namely they use the same coupling function $\mathbf{h}$, but, in general, have different weights $w'_{ij}$. For convenience, we rewrite the input terms by normalizing the weights $w'_{ij}$ by the coupling coefficient $d$, namely $w'_{ij}/d$:
\begin{equation}
    \mathbf{u}_i =d \sum_{j=1}^N w_{ij} \left( \mathbf{h}(\mathbf{x}_i) -\mathbf{h}(\mathbf{x}_j)\right)
\end{equation}
\noindent where $w_{ij}$ are selected such that $w_{ij}+a_{ij} \geq 0$.

We indicate with $\mathrm{A}'$ the adjacency matrix of the controlled network, which therefore includes the links of the pristine structure and the ones introduced by the control, such that $\mathrm{A}'=\mathrm{A}+\mathrm{W}$.

\subsection{Methods for the solution to the control problems}
\label{control:methods}

As already introduced in the Main Text, in this work we consider several control problems that are solved by finding a set of links achieving the specific goal considered, or equivalently finding a suitable matrix $\mathrm{W}$. In general, multiple choices of these links can be performed. For this reason, we look for solution satisfying an optimization problem. In more detail, three different optimization objectives can be considered: minimizing $\| \mathrm{W} \|_2$, preserving the connectedness of the network, and maximizing the sparsity of the solution. We now briefly discuss the three techniques.

\textit{Minimizing the $L_2$ norm of the solution.} This method is based on solving the following optimization problem:
\begin{equation}
    {\rm min} \text{ }||\hat{\mathbf{y}}||_2 \text{ subject to: } \begin{dcases}
       \mathbf{z}+ \hat{\mathbf{y}} \geq 0 \\
        \mathcal{C}_1 \hat{\mathbf{y}} = \mathbf{q}_1 \\
        \mathcal{C}_2 \hat{\mathbf{y}} \geq \mathbf{q}_2
    \end{dcases}
    \label{eq: algorithm_1}
\end{equation}
\noindent where $\hat{\mathbf{y}}$, $\mathbf{z}$, $\mathcal{C}_1$ and $\mathcal{C}_2$ depend on the specific control problem under consideration. This problem can be solved through linear programming  and allows to find the solution that requires the least changes in the interaction network, as measured by the $L_2$ norm of the solution matrix.

\textit{Preserving network connectedness.} This technique preserves the connectivity of the graph and is based on the following optimization problem:
\begin{equation}
    {\rm min} \text{ }||\hat{\mathbf{y}}||_2 \text{ subject to: } \begin{dcases}
        \hat{\mathbf{y}} \geq 0 \\
        \mathcal{C}_1 \hat{\mathbf{y}} = \mathbf{q}_1 \\
        \mathcal{C}_2 \hat{\mathbf{y}} \geq \mathbf{q}_2
    \end{dcases}
        \label{eq: algorithm_2}
\end{equation}

\noindent where, also in this case, $\hat{\mathbf{y}}$, $\mathbf{z}$, $\mathcal{C}_1$ and $\mathcal{C}_2$ are tailored on the specific control problem under consideration. The optimization problem can be solved through linear programming, allowing to find an optimal solution for which the network remains connected.

\textit{Maximizing the sparsity of the solution.}
This approach consists of adding/removing unweighted links to the network and is described by the following optimization problem:
\begin{equation}
    {\rm min} \text{ } ||\hat{\mathbf{y}}||_1 \text{ subject to: } \begin{dcases}
        \hat{\mathbf{y}} + \mathbf{z} \geq 0\\
        \hat{\mathbf{y}} \in \mathbb{Z} \\
        \mathcal{C}_1 \hat{\mathbf{y}} = \mathbf{q}_1 \\
        \mathcal{C}_2 \mathbf{\hat{\mathbf{y}}} \geq \mathbf{q}_2 \\
    \end{dcases}
            \label{eq: algorithm_3}
\end{equation}
\noindent This algorithm allows to find the solution by adding/removing the minimum number of control links and can be solved using integer linear programming.


\subsection{Creation of spectral blocks}
\label{sec:creationspectralblocks}

Consider the multi-agent system in Eq.~\eqref{eq: dynamics} and suppose that the original network of interaction has no spectral blocks. Given a number of $M$ arbitrary groups of nodes, indicated as $C_1, C_2, \ldots, C_M$, we want the nodes in these clusters to form spectral blocks in the controlled multi-agent system, that is, we want to control the interaction network so that it has the spectral blocks $\mathcal{S}_1, \ldots, \mathcal{S}_M$.

The structural condition that associates nodes $i$ and $j$ with a spectral block ($a_{ik}=a_{jk}$ $\forall k \notin C_l$) is similar to a condition that guarantees that the nodes $i$ and $j$ are symmetric ($a_{ik}=a_{jk} \forall k=1, \ldots, N$), but less strict. Therefore, we can impose the condition of symmetries in a fictitious network where the connections within each cluster are neglected (set to be equal to 0). This fictitious network is described by the adjacency matrix $\mathrm{B}$ which is an $N \times N$ matrix with entries $b_{ij}=0$ if $i \in C_l, j \in C_m, l = m$, $b_{ij}= a_{ij}$ otherwise. Therefore, to accomplish the control goal, we can use the results obtained in \cite{gambuzza2020controlling} and select the entries of $\mathrm{W}$ such that:
\begin{equation}
    \mathrm{R}_i(\mathrm{B}+\mathrm{W})-(\mathrm{B}+\mathrm{W})\mathrm{R}_i=0 \quad \quad \forall i=1,2,\ldots, M
    \label{eq: spectral_block_property}
\end{equation}
where $\mathrm{R}_i$ is the permutation matrix that maps the nodes of the cluster $C_i=\{i_1, i_2, \ldots, i_{N_i}\}$ into $\{i_2, \ldots, i_{N_i}, i_1\}$.
By vectorization, Eq.~\eqref{eq: spectral_block_property} can be rewritten as:
\begin{equation}
    \mathcal{R}_i {\rm vec}(\mathrm{W}) = {\rm vec} (\mathrm{R}_i \mathrm{B}-\mathrm{B} \mathrm{R}_i)
\end{equation}
\noindent with $\mathcal{R}_i = \mathrm{I}_N \otimes \mathrm{R}_i - \mathrm{R}^T_i \otimes \mathrm{I}_N$. To simultaneously satisfy the condition $\forall i = 1, \ldots, M$, we consider the following equation:
\begin{equation}
    \mathcal{R} {\rm vec}(\mathrm{W}) = \mathbf{b}
    \label{eq: controlloI}
\end{equation}
\noindent where $\mathcal{R}=\begin{bmatrix}
    \mathcal{R}_1^T & \mathcal{R}_2^T \ldots \mathcal{R}_M^T
\end{bmatrix}^T
$ and $\mathbf{b}=\begin{bmatrix}
    ({\rm vec}(\mathrm{R}_1 \mathrm{B}-\mathrm{B} \mathrm{R}_1))^T & ({\rm vec}(\mathrm{R}_2 \mathrm{B}-\mathrm{B} \mathrm{R}_2))^T & \ldots & ({\rm vec}(\mathrm{R}_M \mathrm{B}-\mathrm{B} \mathrm{R}_M))^T
\end{bmatrix}^T$.
As shown in \cite{gambuzza2020controlling}, a solution to \eqref{eq: controlloI} always exists. Indeed, it is sufficient to ensure that $a_{ik}+w_{ik}=a_{jk}+w_{jk}=1$ $\forall i,j \in C_l$ and $\forall k \notin C_l$ $\forall l=1, \ldots, M$. However, this solution is inefficient since it requires adding many control links.
To find the optimal solution, one of the three algorithms discussed in Sec.~\ref{control:methods} can be used, with $\hat{\mathbf{y}}={\rm vec}(\mathrm{W})$, $\mathbf{z}={\rm vec}(\mathrm{A})$, $\mathcal{C}_1=\mathcal{R}$, $\mathbf{q}_1=\mathbf{b}$, $\mathcal{C}_2 = 0$, and $\mathbf{q}_2 = \mathbf{0}$.

\newpage
\subsection{Controlling cluster synchronizability}
\label{sec: control cluster synchronizability}
We consider again the multi-agent system described by Eq.~\eqref{eq: dynamics}, but we now suppose that the interaction network already has $M$ spectral blocks $\mathcal{S}_1, \mathcal{S}_2, \ldots, \mathcal{S}_M$. Assume that the system has a class III MSF and that some of these spectral blocks are not synchronizable because they do not satisfy the condition on the eigenvalue ratio. In other terms, taking into account the properties of synchronization of the spectral blocks as discussed in the Main Text, the $M'\leq M$ subgraphs $\mathcal{G}_1, \mathcal{G}_2, \ldots, \mathcal{G}_{M'}$ associated with $\mathcal{S}_1, \mathcal{S}_2, \ldots, \mathcal{S}_{M'}$ are such that:
\begin{equation}
    \frac{\lambda_{N_l}(\mathcal{L}_l)+s_l}{\lambda_{2}(\mathcal{L}_l)+s_l}\geq \frac{\nu_2^*}{\nu_1^*} \quad \quad \forall l=1,\ldots,M'
\label{eq: condizione_precontrollo2}
\end{equation}
\noindent where $N_l=|C_l|$, $\mathcal{L}_l$ is the Laplacian matrix of $\mathcal{G}_l$, $s_l$ is the strength through which each node of the cluster $C_l$ associated with $\mathcal{S}_l$ is connected to the rest of the graph, i.e., $s_l= \sum\limits_{j \notin  C_l} a_{ij}$ for $i \in C_l$, and $\nu_1^*$ and $\nu_2^*$ are the two critical points characterizing the system MSF. Because of Eq.~\eqref{eq: condizione_precontrollo2}, there are no values of $d$ for which the clusters can synchronize. The goal of the control is to let such clusters become synchronizable.

To this aim, we select the weights of the control links to satifsy two conditions. First, we want to preserve the structural condition of the existence of the original clusters, that is, the clusters $C_1, C_2, \ldots, C_M$ must still to be associated with spectral blocks. The, we require that they now satisfy the condition on the eigenvalue ratio. In practice, the control input has to be selected such that $\mathrm{W}=\mathrm{W}^T$, $w_{ik}=w_{jk}  \forall i,j \in C_l, \forall k \notin C_l, \forall l=1,\ldots, M$, and:
\begin{equation}
     \frac{\lambda_{N_l}(\mathcal{L}_l)+s_l+w_l}{\lambda_2(\mathcal{L}_l)+s_l+w_l}<\frac{\nu_2^*}{\nu_1^*} \quad \quad \forall l=1,\ldots, M'
\label{eq: controlloII_1}
\end{equation}

\noindent with $w_l$ being defined as $w_l = \sum\limits_{j\notin C_l} w_{ij}$ for $i \in C_l$.
By rearranging the terms of \eqref{eq: controlloII_1}, we obtain:
\begin{equation}
          w_l+s_l-  \left( \frac{\nu_2^*}{\nu_1^*} -1\right)^{-1} \left( \lambda_{N_l}(\mathcal{L}_l)- \frac{\nu_2^*}{\nu_1^*} \lambda_2(\mathcal{L}_l) \right)>0, \quad \quad \forall l=1,\ldots, M'
    \label{eq: controlloII_2}
\end{equation}

Next we consider the quotient graph $\mathcal{G}/\pi$ and its associated adjacency matrix $\mathrm{S}$, and we introduce the matrix $\mathrm{X} \in \mathbb{R}^{N_\pi \times N_\pi}$ whose elements are $x_{lm}= w_{ij}$ $\forall i \in C_l$, $\forall j \in C_m$, $l \neq m$.
Let $\mathrm{O}_{K, \Omega}=\begin{bmatrix}
    \mathrm{I}_{K} & \mathrm{0}_{K, \Omega-K}
\end{bmatrix}$ be a $K \times \Omega$ matrix such that the product $\mathrm{O}_{K,\Omega} Z$ returns the first $K$ rows of $Z$. Then, Eqs.~\eqref{eq: controlloII_2} can be written in compact, matrix-vector form as follows:
\begin{equation}
    \begin{dcases}
         \mathrm{O}_{M', N_\pi} \left(\mathrm{X}+\mathrm{S}\right) \mathbf{\gamma}-\left( \frac{\nu_2^*}{\nu_1^*}-1 \right)^{-1}  \mathrm{O}_{M',M} \left( \mathbf{\Lambda_N}- \frac{\nu_2^*}{\nu_1^*} \mathbf{\Lambda_2}  \right) > \mathbf{0} \\
         {\mathrm{X}}-{\mathrm{X}}^T=0 \\
         {\rm diag}(\mathrm{X})=0 \\
    \end{dcases}
    \label{eq:controlloII_matrix}
\end{equation}

\noindent where $\mathbf{\gamma}=[N_1, N_2, \ldots, N_{N_\pi}]^T$, $\mathbf{\Lambda_2}=[\lambda_2(\mathcal{L}_1), \lambda_2(\mathcal{L}_2), \ldots, \lambda_2(\mathcal{L}_M)]^T$,
$\mathbf{\Lambda_N}=[\lambda_{N_1}(\mathcal{L}_1), \lambda_{N_2}(\mathcal{L}_2), \ldots, \lambda_{N_M}(\mathcal{L}_M)]^T$ and $\mathbf{e_l} \in \mathbb{R}^{
N_\pi}$ is the $l$-th canonical vector.

By vectorization, Eq.~\eqref{eq:controlloII_matrix} is rewritten as:
\begin{equation}
\begin{dcases}
        \mathrm{F} \mathbf{x} > \mathbf{g} \\
    \mathrm{E} \mathbf{x} = \mathbf{0}
\end{dcases}
\label{eq: controlloII_vect}
\end{equation}
\noindent where $\mathbf{x}={\rm vec} \left(\mathrm{X} \right)$, $\mathrm{F}=\mathbf{\gamma}^T \otimes \mathrm{O}_{M', N_\pi}$,
         $\mathbf{g}= -\mathrm{O}_{M', N_\pi} \mathrm{S} \mathbf{\gamma} + \left( \frac{\nu_2^*}{\nu_1^*}-1 \right)^{-1} \mathrm{O}_{M',M}\left( \mathbf{\Lambda_N}-\frac{\nu_2^*}{\nu_1^*}\mathbf{\Lambda_2}  \right)$.
 and $\mathrm{E}=\begin{bmatrix}
     \mathrm{E}_1 \\ \mathrm{E}_2
 \end{bmatrix}$ is a block matrix with:

 \begin{equation}
 \label{eq:cp2tosolve}
    \mathrm{E_1}=\mathrm{I}_{N_\pi}-\begin{bmatrix}
        \mathrm{I}_{N_\pi} \otimes \mathbf{e_1} \\
        \mathrm{I}_{N_\pi} \otimes \mathbf{e_2} \\
        \vdots \\
        \mathrm{I}_{N_\pi} \otimes \mathbf{e_{N_\pi}}
    \end{bmatrix},
            \quad \quad
      \mathrm{E_2}=\begin{bmatrix}
        \mathbf{e_1} &  & &\\
        & \mathbf{e_2} &  &\\
        & & \ddots & \\
        & & & \mathbf{e_{N_\pi}}\end{bmatrix}
 \end{equation}
 \noindent where the matrix $\mathrm{E}_1$ is used to set $\mathrm{X}=\mathrm{X}^T$, while $\mathrm{E}_2$ to set ${\rm diag}(\mathrm{X})=0$.

We solve~\eqref{eq:cp2tosolve} for the unknowns $x_{lm}$ and then, once we have obtained them, we compute the entries of $W$ as follows:
\begin{equation}
    w_{ij}=x_{lm} \quad \quad \forall i\in C_l,  \forall j \in C_m, \forall l,m=1, \ldots ,N_\pi
    \label{eq: entries_W}
\end{equation}

Notice that a solution of Eq.~\eqref{eq: controlloII_vect} always exists since \begin{equation}
\lim_{w_l\to\infty} \frac{\lambda_{N_l}(\mathcal{L}_l)+s_l+w_l}{\lambda_2(\mathcal{L}_l)+s_l+w_l}=1<\frac{\nu_2^*}{\nu_1^*} \quad \quad \forall l=1,\ldots,M
\end{equation}

To determine the optimal solution, one of the three algorithms described in Sec.~\ref{control:methods} can be used, where $\hat{\mathbf{y}}={\rm vec}(\mathrm{X})$, $\mathbf{z}={\rm vec}(\mathrm{S})$, $\mathcal{C}_1= E$, $\mathbf{q}_1=\mathbf{0}$, $\mathcal{C}_2=\mathrm{F}$, $\mathbf{q}_2=\mathbf{g}+\epsilon \mathbf{1}$, where $\epsilon>0$ is an arbitrary small number. It is worth noticing that $\mathbf{q}_2$ is not equal to $\mathbf{g}$ because the inequality in Eq.~\eqref{eq: controlloII_vect} is strict, unlike those in the optimization problems in Sec.~\ref{control:methods}.

\begin{figure}[H]
\centering
\subfigure[]{\includegraphics[width=0.4\textwidth]{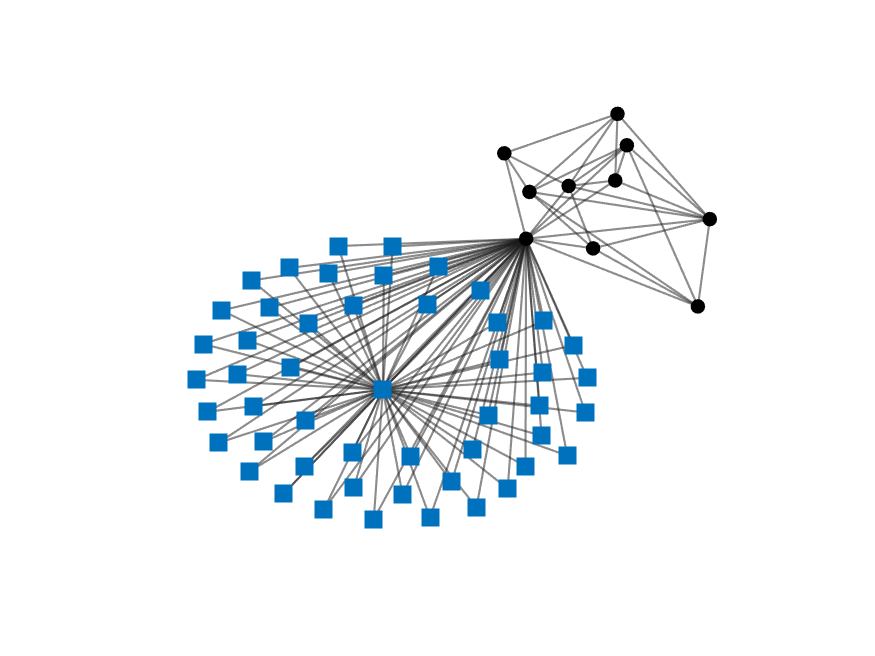}
\label{fig:grafo_controllo2_sm_iniziale}}
\subfigure[]{\includegraphics[width=0.4\textwidth]{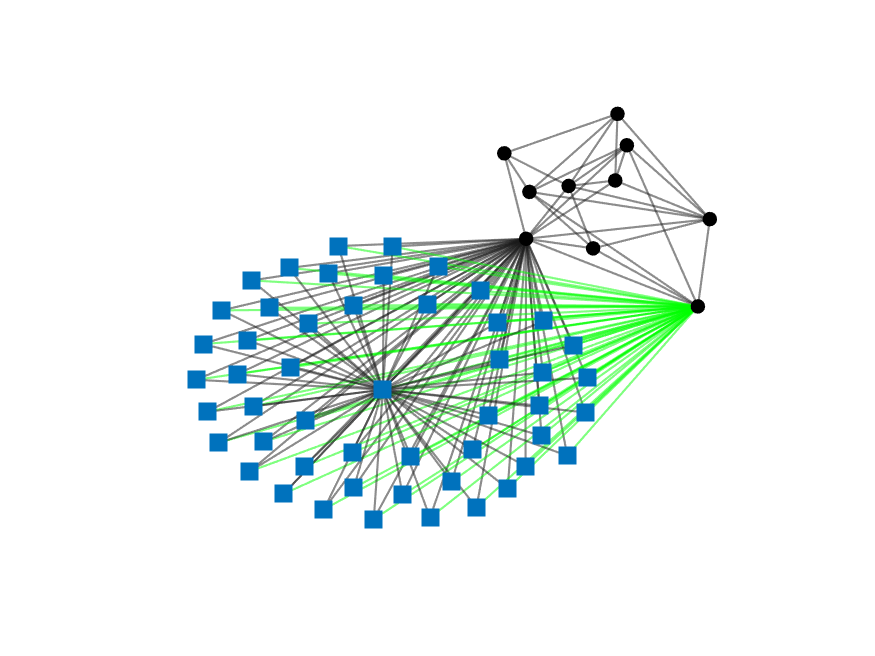}
\label{fig:grafo_controllo2_sm}}
\subfigure[]{\includegraphics[width=0.4\textwidth]{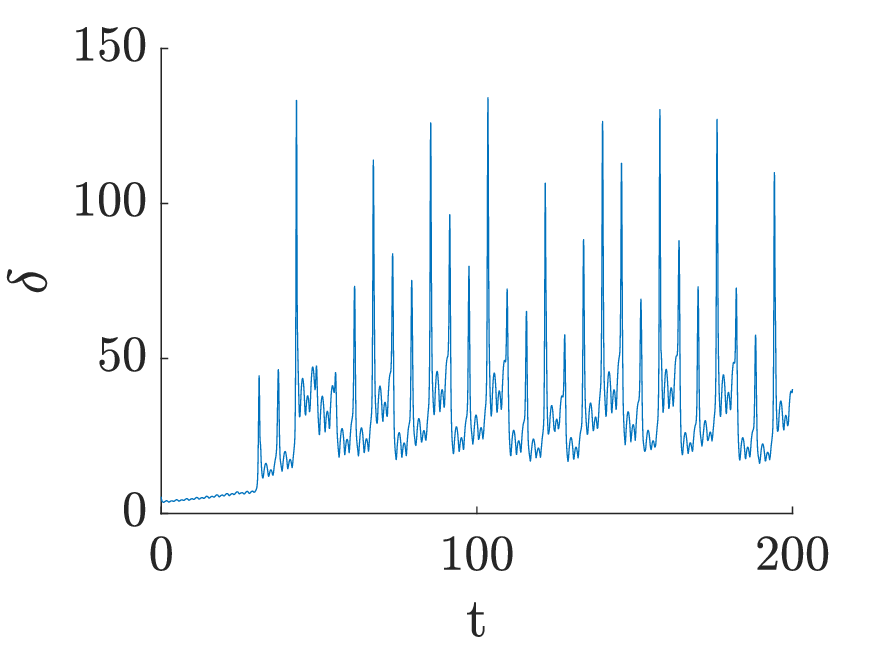}
\label{fig:traiettoria_sm1}}
\subfigure[]{\includegraphics[width=0.4\textwidth]{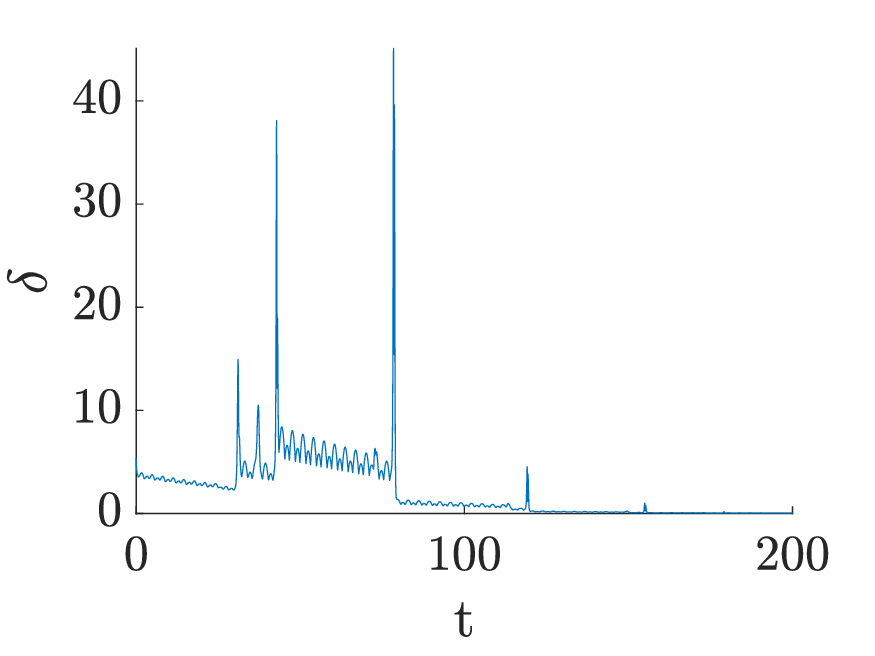}
\label{fig:traiettoria_sm2}}

\caption{(a) Pristine network equipped with a cluster $C'$ associated with a spectral block $\mathcal{S}$. The nodes belonging to the cluster are connected to the bulk with a strength $s'=1$. (b) Controlled network with the nodes belonging to the cluster associated with $S$ being connected to the bulk with a strength $s'+w=2$, the added links are represented in green. (c) Time evolution of the synchronization error $\delta(t)=$ in the absence of control with $d=0.075$. (d) Time evolution of the synchronization error $\delta(t)= \frac{1}{N'} \left( \sum_{i \in C'} ||\mathbf{x}_i(t)-\bar{\mathbf{x}}(t)||^2 \right) ^{\frac{1}{2}}$ when the network is controlled with $d=0.075$.}
\end{figure}

In the main text, we showed an example where each unit was associated with Lorenz dynamics; here, we examine another example involving Rössler dynamics.
Specifically, we consider the unweighted network of $N=60$ nodes shown in Fig.~\ref{fig:grafo_controllo2_sm_iniziale} that is equipped with a spectral block $\mathcal{S}$ localized at nodes $C'=\{1, \ldots, 50\}$ (these nodes are represented by blue squares).

Each node of the multi-agent system is a R\"ossler oscillator with dynamics described by the following equation:
\begin{equation}
\begin{cases}
\dot{x}_{i,1}=-{x}_{i,2}-{x}_{i,3}+d \sum\limits_{j=1}^N \left({a}_{ij}+w_{ij} \right) \left({x}_{j,1}-{x}_{i,1} \right)\\
\dot{x}_{i,2}={x}_{i,1}+a{x}_{i,2}\\
\dot{x}_{i,3}=b+{x}_{i,3}({x}_{i,1}-c)\\
\end{cases}
\label{eq:RosslerCasoMSFTipoIII}
\end{equation}
with $a=b=0.2$, and $c=7$. This system has a type III MSF with $\nu_1^*=0.186$ and $\nu_2^*=4.614$ \cite{huang2009generic}.

Let $\mathcal{L}'$ be the Laplacian matrix of the subgraph $\mathcal{G}'$ associated with $C'$, the smallest and the largest non-zero eigenvalues of $\mathcal{L}'$ are $\lambda_2(\mathcal{L}')=1$ and $\lambda_{50}(\mathcal{L}')=50$. Furthermore, the cluster $C'$ is connected to the rest of the graph with a strength $s_1=1$. Since $\frac{\lambda_{50}(\mathcal{L}')+s'}{\lambda_{2}(\mathcal{L}')+s'}>\frac{\nu_2^*}{\nu_1^*}$, the nodes of cluster $C'$ cannot synchronize for any value of $d$. This can be also observed in Fig.~\ref{fig:traiettoria_sm1} that shows the temporal evolution of the synchronization error $\delta(t) =\frac{1}{N'} \left( \sum_{i \in C'} ||\mathbf{x}_i(t)-\bar{\mathbf{x}}(t)||^2 \right) ^{\frac{1}{2}}$ (with $N' = |C'|$ and $\bar{\mathbf{x}}(t)=\frac{1}{N'}\sum_{j \in C'} \mathbf{x}_j(t)$) for $d=0.075$.

The control action is implemented using the method based on maximizing the sparsity of the solution to determine the optimal solution for Eq.~\eqref{eq: controlloII_1}. Fig.~\ref{fig:grafo_controllo2_sm} shows the controlled network, in which $C'$ is connected to the bulk with a strength $s_1+w_1=2$, where the added links are represented in green. Since $\frac{\lambda_{50}(\mathcal{L}_1)+s_1+w_1}{\lambda_{2}(\mathcal{L}_1)+s_1+w_1}<\frac{\nu_2^*}{\nu_1^*}$, then $C'$ satisfies the condition on the eigenvalue ratio and does synchronize for $0.062=\frac{\nu_1^*}{\lambda_2(\mathcal{L}')+s'+w'}<d<\frac{\nu_2^*}{\lambda_{50}(\mathcal{L}')+s'+w'}=0.088$. By selecting $d=0.075$, the cluster $C'$ reaches synchronization and, as shown in Fig.~\ref{fig:traiettoria_sm2}, the synchronization error $\delta(t)$ vanishes in time.

\subsection{Shaping the synchronization/desynchronization sequence}

Here, we consider a multi-agent system described by Eq.~\eqref{eq: dynamics} where the interaction graph is endowed with $M$ spectral blocks, $\mathcal{S}_1, \mathcal{S}_2, \ldots, \mathcal{S}_M$. The clusters $C_1, C_2, \ldots, C_M$ associated with $\mathcal{S}_1, \mathcal{S}_2, \ldots, \mathcal{S}_M$ are all synchronizable, but we want to control the sequence in which they synchronize and desynchronize as $d$ increases. Eventually, this multi-agent system may be the result of the application of the two control methods described in Secs.~\ref{sec:creationspectralblocks} and~\ref{sec: control cluster synchronizability}.

Assuming that the desired sequences are $\mathcal{I}'=\mathcal{D}'=\{1,2,\ldots, M\}$, the weights of the control links have to be selected such that three conditions are satisfied: \textit{i)} the structural conditions for the spectral blocks $S_1, S_2, \ldots, S_M$, to which the clusters $C_1, C_2, \ldots, C_M$ are associated, are preserved after the inclusion of the control links in the network of interaction; \textit{ii)} the constraint on the eigenvalue ratio is satisfied for each cluster in the controlled network; \textit{iii)} the synchronization and desynchronization sequence are equal to $\mathcal{I}'$ and $\mathcal{D}'$, respectively. In practice, the entries of $\mathrm{W}$ are selected such that Eqs.~\eqref{eq: controlloII_1} hold, $\mathrm{W}=\mathrm{W}^T$, $w_{ik}=w_{jk}$  $\forall i,j \in C_l, \forall k \notin C_l, \forall l=1,\ldots, M$, and:
\begin{equation}
\begin{dcases}
        \frac{\nu_1^*}{\lambda_2(\mathcal{L}_l)+s_l+w_l} < \frac{\nu_1^*}{\lambda_2(\mathcal{L}_{l+1})+s_{l+1}+w_{l+1}} & \forall l=1,\ldots, M-1 \\
        \frac{\nu_2^*}{\lambda_{N_l}(\mathcal{L}_l)+s_l+w_l} < \frac{\nu_2^*}{\lambda_{N_{l+1}}(\mathcal{L}_{l+1})+s_{l+1}+w_{l+1}} & \forall l=1,\ldots, M-1 \\
        \label{eq: controlloIII}
\end{dcases}
\end{equation}

By rearranging the terms in Eq.~\eqref{eq: controlloIII}, we obtain that:
\begin{equation}
\begin{dcases}
        s_l + w_l - s_{l+1} - w_{l+1} + \lambda_2(\mathcal{L}_l)-\lambda_2(\mathcal{L}_{l+1})>0 & \forall l=1,\ldots, M-1\\
        s_l + w_l - s_{l+1} - w_{l+1} + \lambda_{N_l}(\mathcal{L}_l)-\lambda_{N_{l+1}}(\mathcal{L}_{l+1})>0 & \forall l=1,\ldots, M-1
\end{dcases}
\label{eq: controlloIII_2}
\end{equation}

We now proceed as we did for the control problem illustrated in Sec.~\ref{sec: control cluster synchronizability}, rewriting Eqs.~\eqref{eq: controlloIII_2} in compact form via the quotient graph $\mathcal{G}/\pi$, its adjacency matrix $\mathrm{S}$ and the matrix $\mathrm{X}$, as follows:
\begin{equation}
\begin{dcases}
         \left( \mathbf{1}_2 \otimes \mathrm{Q} \right) \mathrm{O}_{M, N_\pi} \left( \mathrm{S}+\mathrm{X} \right) \mathbf{\gamma}+\left( \mathrm{I}_2 \otimes \mathrm{Q} \right) \begin{bmatrix}
         \mathbf{\Lambda_2}  \\
         \mathbf{\Lambda_N}
         \end{bmatrix} > \mathbf{0} \\
         {\mathrm{X}}-{\mathrm{X}}^T=0 \\
         {\rm diag}(\mathrm{X}) = 0
\end{dcases}
         \label{eq: controlloIII_matrix}
\end{equation}

\noindent where
\begin{equation}
        \mathrm{Q}=\begin{bmatrix}
        1 & -1 & \\
        & 1 & -1 & \\
        & & \ddots & \ddots \\
        & & & 1 & -1 \\
     \end{bmatrix}=\begin{bmatrix}
        \mathrm{I}_{M-1} & \mathbf{0}_{M-1}
    \end{bmatrix} - \begin{bmatrix}
        \mathbf{0}_{M-1} & \mathrm{I}_{M-1}
    \end{bmatrix}
\end{equation}
We note that the matrix $\mathrm{Q} \in \mathbb{R}^{(M-1) \times M}$ has the following property. Given a $M \times K$ matrix $\mathrm{Z}$, then the $i-$th row of the product $\mathrm{QZ}$ is $(\mathrm{Q}\mathrm{Z})_i=\mathrm{Z}_i-\mathrm{Z}_{i+1}$, where $\mathrm{Z}_i$ denotes the $i-$th row of $\mathrm{Z}$.

By performing the vectorization of Eq.~\eqref{eq: controlloIII_matrix} and by considering Eq.~\eqref{eq: controlloII_vect}, which is the vectorization of the condition \textit{ii}), we obtain the following:
\begin{equation}
\begin{dcases}
    \mathrm{F} \mathbf{x} > \mathbf{g} \\
    \mathrm{P} \mathbf{x} > \mathbf{c} \\
    \mathrm{E} {\mathbf{x}} = \mathbf{0}
\end{dcases}
\label{eq:controlloIII_vector_sm}
\end{equation}

\noindent where $\mathrm{P}=\left( \mathbf{\gamma}^T \otimes \mathbf{1}_2 \otimes \mathrm{Q} \mathrm{O}_{M, N_\pi} \right)$, and $\mathbf{c}=- \left(\mathbf{1}_2 \otimes \mathrm{Q}\right) \mathrm{O}_{M, N_\pi} \mathrm{S} \mathbf{\gamma} -  \left(\mathrm{I}_2 \otimes \mathrm{Q}\right) \begin{bmatrix}
         \mathbf{\Lambda_2}  \\
         \mathbf{\Lambda_N}
         \end{bmatrix}$.

Once the entries of $\mathrm{X}$ are obtained, we compute the coefficients $\mathrm{w}_{ij}$ using Eq.~\eqref{eq: entries_W}.


Here we do not provide a formal proof of the existence of a solution $\mathbf{x}$ of Eq.~\eqref{eq:controlloIII_vector_sm}, such that $\mathbf{x}  +{\rm vec}(\mathrm{S}) \geq 0$, but only sketch the arguments behind it. Given the vector $\hat{\mathbf{c}}=\mathbf{c}+(\mathrm{1}_2 \otimes \mathrm{Q}) \mathrm{O}_{M,N_\pi} \mathrm{S} \mathbf{\gamma}$, the second inequality of Eq.~\eqref{eq:controlloIII_vector_sm} is equivalent to $\mathrm{P} \left(\mathbf{x}+ \rm vec(\mathrm{S}) \right)>\hat{\mathbf{c}}$. 
Now, let us consider the matrix $\mathrm{G}=\mathrm{P}\mathrm{T}$ where $\mathrm{T}$ is defined as 
\begin{equation}
    \mathrm{T}=\mathrm{I}_{N_\pi}+\begin{bmatrix}
        \mathrm{I}_{N_\pi} \otimes \mathbf{e_1} \\
        \mathrm{I}_{N_\pi} \otimes \mathbf{e_2} \\
        \vdots \\
        \mathrm{I}_{N_\pi} \otimes \mathbf{e_{N_\pi}} 
    \end{bmatrix}
\end{equation}
\noindent and let us then remove from $\mathrm{G}$ the columns that correspond to the elements $x_{lm}$ with $l \geq m$, thus obtaining another matrix that we indicate as $\hat{\mathrm{G}}$. By construction, the matrix $\hat{\mathrm{G}}$ is such that there is no $\mathbf{y} \geq \mathbf{0}$ that satisfies $-\hat{\mathrm{G}}^T \mathbf{y} \geq 0$ and $-\hat{\mathbf{c}}^T \mathbf{y}>0$. Therefore, for Lemma~\ref{lemma: Farkas}, Eq.~\eqref{eq:controlloIII_vector_sm} always has a solution $\mathbf{x} \geq -\rm vec(\mathrm{S})$. Finally, we can replace $\hat{\mathbf{c}}$ with $\hat{\mathbf{c}}+\epsilon \mathbf{1}$ to obtain a solution strictly positive.


Analogously to the control problems discussed above, the solution of Eq.~\eqref{eq:controlloIII_vector_sm} can be obtained through one of the three approaches shown in Sec.~\ref{control:methods}, with $\hat{\mathbf{y}}={\rm vec}(\mathrm{X})$, $\mathbf{z}={\rm vec}(\mathrm{S})$, $\mathcal{C}_1= \mathrm{E}$, $\mathbf{q}_1=\mathbf{0}$, $\mathcal{C}_2=\begin{bmatrix}
    \mathrm{F} \\
     \mathrm{P}
\end{bmatrix}$, $\mathbf{q}_2= \begin{bmatrix}
    \mathbf{g} \\
    \mathbf{c}
\end{bmatrix}+\epsilon \mathbf{1}$.

\begin{figure}[H]
\centering
\subfigure[]{\includegraphics[width=0.33\textwidth]{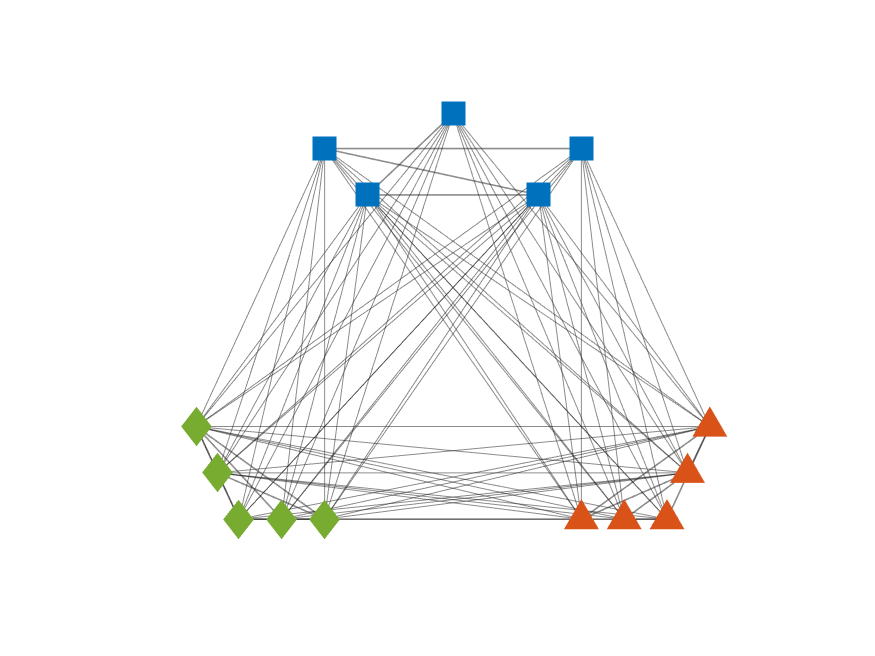}
\label{fig:grafo_controllo3_sm_iniziale}}
\subfigure[]{\includegraphics[width=0.33\textwidth]{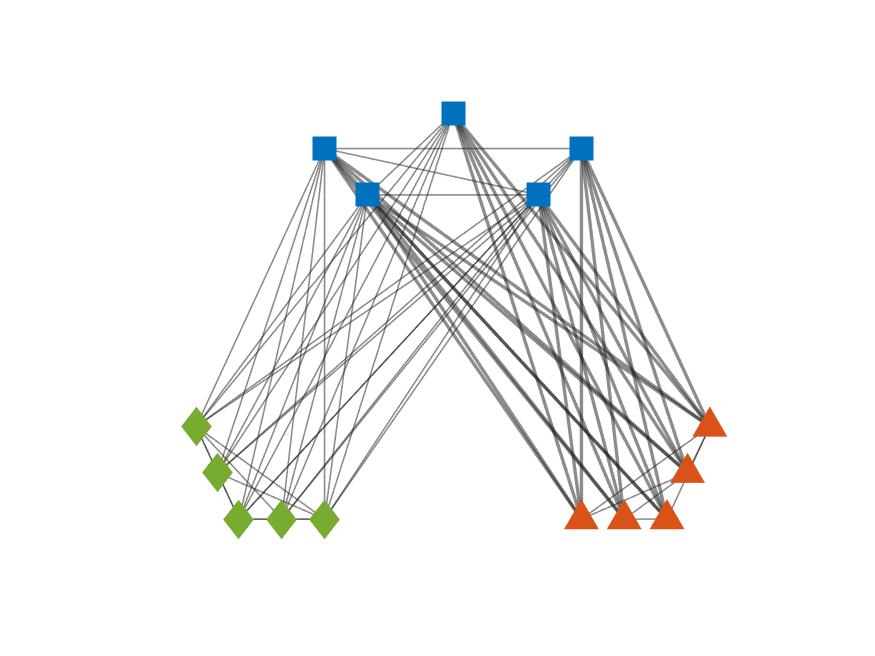}
\label{fig:grafo_controllo3_sm}}
\subfigure[]{\includegraphics[width=0.32\textwidth]{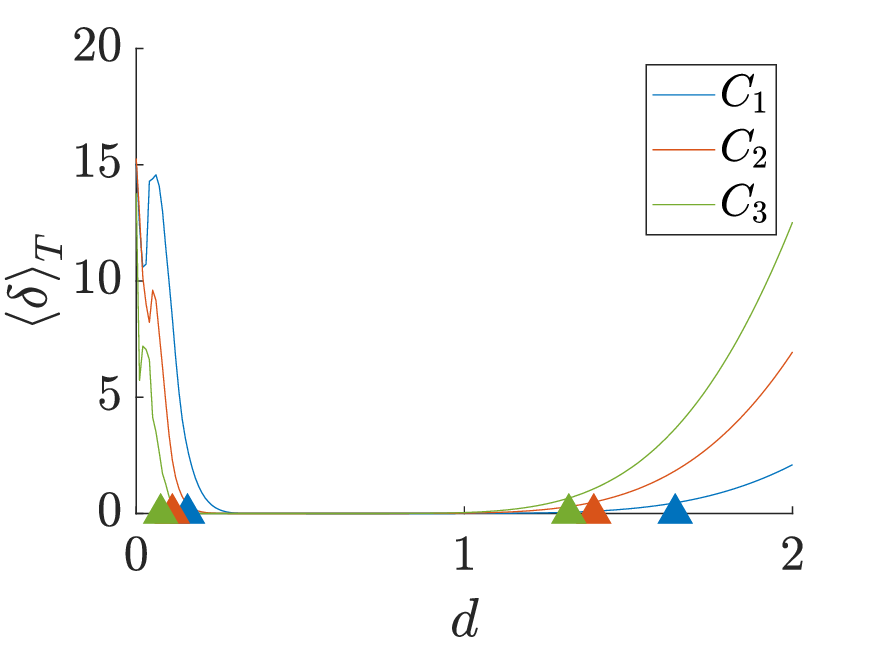}
\label{fig:errore_sm1}}
\subfigure[]{\includegraphics[width=0.32\textwidth]{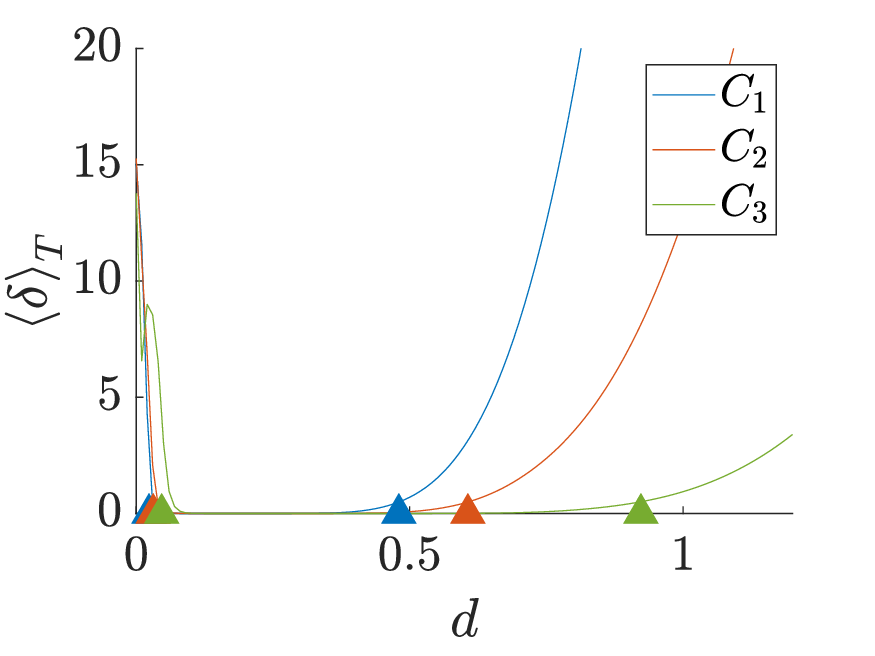}
\label{fig:errore_sm2}}
\caption{(a) Pristine weighted network, where the thickness of each edge is proportional to the edge weight. The network is formed by three spectral blocks $\mathcal{S}_1, \mathcal{S_2}, \mathcal{S}_3$ localized at the nodes of the clusters $C_1$ (blue squares), $C_2$ (orange triangles) and $C_3$ (green diamonds). These clusters are connected with the rest of the graph with strength $s_1=s_2=s_3=1$. (b) Controlled network, where the clusters are connected with the rest of the graph with strength $s_1+w_1=7.8$, $s_2+w_2=5.3$, and $s_3+w_3=2.5$, respectively. (c,d) $\langle \delta_h \rangle_T$  vs. $d$ for the uncontrolled (panel c) and controlled (panel d) network for the three clusters $C_1$ (blue line), $C_2$ (orange line) and $C_3$ (green line). The triangles mark the transition values for synchronization stability predicted by the MSF approach. Controlling the multi-agent system makes possible to change the synchronization and desynchronization sequence from $\mathcal{I}=\mathcal{D}=\{3,2,1\}$ to  $\mathcal{I'}=\mathcal{D'}=\{1,2,3\}$.}
\end{figure}

Let us now discuss an example where the optimal solution is obtained using the method based on minimizing the $L_2$ norm, as in Sec.~\ref{control:methods}. We consider the weighted interaction network shown in Fig.~\ref{fig:grafo_controllo3_sm_iniziale}, where the width of each link is proportional to its weight. This network is composed of three spectral blocks, $\mathcal{S}_1$, $\mathcal{S}_2$, and $\mathcal{S}_3$, localized at the nodes of the clusters $C_1=\{1,\ldots,5\}$, $C_2=\{6,\ldots,10\}$, and $C_3=\{11,\ldots,15\}$. The units of the network are R\"ossler oscillators, described by Eq.~\eqref{eq:RosslerCasoMSFTipoIII}. As mentioned above, this system has a type III MSF with $\nu_1^*=0.186$ and $\nu_2^*=4.614$. The adjacency matrix associated with the multi-agent system (in absence of control) is:

\setcounter{MaxMatrixCols}{20}
\begin{equation*}
\mathrm{A}=\begin{bmatrix}
0 & 0.5 & 0 & 0 & 0 & 0.1 & 0.1 & 0.1 & 0.1 & 0.1 & 0.1 & 0.1 & 0.1 & 0.1 & 0.1 \\
0.5 & 0 & 0.5 & 0 & 0 & 0.1 & 0.1 & 0.1 & 0.1 & 0.1 & 0.1 & 0.1 & 0.1 & 0.1 & 0.1 \\
0 & 0.5 & 0 & 0.5 & 0 & 0.1 & 0.1 & 0.1 & 0.1 & 0.1 & 0.1 & 0.1 & 0.1 & 0.1 & 0.1 \\
0 & 0 & 0.5 & 0 & 0.5 & 0.1 & 0.1 & 0.1 & 0.1 & 0.1 & 0.1 & 0.1 & 0.1 & 0.1 & 0.1 \\
0 & 0 & 0 & 0.5 & 0 & 0.1 & 0.1 & 0.1 & 0.1 & 0.1 & 0.1 & 0.1 & 0.1 & 0.1 & 0.1 \\
0.1 & 0.1 & 0.1 & 0.1 & 0.1 & 0 & 0.5 & 0.5 & 0.5 & 0 & 0.1 & 0.1 & 0.1 & 0.1 & 0.1 \\
0.1 & 0.1 & 0.1 & 0.1 & 0.1 & 0.5 & 0 & 0 & 0 & 0.5 & 0.1 & 0.1 & 0.1 & 0.1 & 0.1 \\
0.1 & 0.1 & 0.1 & 0.1 & 0.1 & 0.5 & 0 & 0 & 0.5 & 0.5 & 0.1 & 0.1 & 0.1 & 0.1 & 0.1 \\
0.1 & 0.1 & 0.1 & 0.1 & 0.1 & 0.5 & 0 & 0.5 & 0 & 0 & 0.1 & 0.1 & 0.1 & 0.1 & 0.1 \\
0.1 & 0.1 & 0.1 & 0.1 & 0.1 & 0 & 0.5 & 0.5 & 0 & 0 & 0.1 & 0.1 & 0.1 & 0.1 & 0.1 \\
0.1 & 0.1 & 0.1 & 0.1 & 0.1 & 0.1 & 0.1 & 0.1 & 0.1 & 0.1 & 0 & 0.5 & 0 & 0.5 & 0.5 \\
0.1 & 0.1 & 0.1 & 0.1 & 0.1 & 0.1 & 0.1 & 0.1 & 0.1 & 0.1 & 0.5 & 0 & 0.5 & 0.5 & 0.5 \\
0.1 & 0.1 & 0.1 & 0.1 & 0.1 & 0.1 & 0.1 & 0.1 & 0.1 & 0.1 & 0 & 0.5 & 0 & 0.5 & 0.5 \\
0.1 & 0.1 & 0.1 & 0.1 & 0.1 & 0.1 & 0.1 & 0.1 & 0.1 & 0.1 & 0.5 & 0.5 & 0.5 & 0 & 0.5 \\
0.1 & 0.1 & 0.1 & 0.1 & 0.1 & 0.1 & 0.1 & 0.1 & 0.1 & 0.1 & 0.5 & 0.5 & 0.5 & 0.5 & 0
\end{bmatrix}
\end{equation*}

The strengths with which the clusters $C_1$, $C_2$, and $C_3$ are connected to the rest of the graph are $s_1=s_2=s_3=1$, since $a_{ij}=0.1$ $\forall i \in C_l, j \in C_m, l\neq m$.
In addition, the smallest and the largest non-zero eigenvalues of $\mathcal{L}_1$, $\mathcal{L}_2$, $\mathcal{L_3}$ are $\lambda_2(\mathcal{L}_1)=0.19$, $\lambda_2(\mathcal{L}_2)=0.69$, $\lambda_2(\mathcal{L}_3)=1.51$, and $\lambda_5(\mathcal{L}_1)=1.81$, $\lambda_5(\mathcal{L}_2)=2.31$, $\lambda_5(\mathcal{L}_3)=2.5$. Therefore, since  $\lambda_2(\mathcal{L}_1)+s_1 < \lambda_2(\mathcal{L}_2)+s_2 < \lambda_2(\mathcal{L}_3)+s_3$ and  $\lambda_5(\mathcal{L}_1)+s_1 < \lambda_5(\mathcal{L}_2)+s_2 < \lambda_5(\mathcal{L}_3)+s_3$, in the absence of control, the synchronization and desynchronization sequences are $\mathcal{I}=\mathcal{D}= \{3, 2, 1\}$. This is confirmed by the numerical simulations of the multi-agent system in Eqs.~\eqref{eq:RosslerCasoMSFTipoIII} illustrated in Fig.~\ref{fig:errore_sm1}. The time evolution of the system variables has been computed for a period of time equal to $5T$, with $T=10$. After discarding a transient of $4T$, we have calculated the average value (on a window of time $T$) of the cluster synchronization error $\langle \delta_h \rangle_T= \langle \frac{1}{N'} \left( \sum_{i \in C_h} ||\mathbf{x}_i-\mathbf{\bar{x}_h}||^2 \right) ^{\frac{1}{2}} \rangle_T$, with $h=\{1,2,3\}$, for each of the three clusters of the network, namely $C_1$ (blue curve), $C_2$ (orange curve) and $C_3$ (green curve), as function of the coupling strength $d$.  Fig.~\ref{fig:errore_sm1} also shows the critical values predicted by the MSF approach marked as blue, orange or green triangles for the three clusters, $C_1$, $C_2$, and $C_3$.

Next, we apply the control of the multi-agent system, using minimization of $||\mathrm{W}||_2$ to find a solution of Eq.~\eqref{eq:controlloIII_vector_sm} with desired synchronization/desynchronization sequence  $\mathcal{I}'=\mathcal{D}'=\{1,2,3\}$.

This leads to the following matrix $\mathrm{W}$:

\begin{equation*}
 \mathrm{W}=   \begin{bmatrix}
0 & 0 & 0 & 0 & 0 & 0.96 & 0.96 & 0.96 & 0.96 & 0.96 & 0.4 & 0.4 & 0.4 & 0.4 & 0.4 \\
0 & 0 & 0 & 0 & 0 & 0.96 & 0.96 & 0.96 & 0.96 & 0.96 & 0.4 & 0.4 & 0.4 & 0.4 & 0.4 \\
0 & 0 & 0 & 0 & 0 & 0.96 & 0.96 & 0.96 & 0.96 & 0.96 & 0.4 & 0.4 & 0.4 & 0.4 & 0.4 \\
0 & 0 & 0 & 0 & 0 & 0.96 & 0.96 & 0.96 & 0.96 & 0.96 & 0.4 & 0.4 & 0.4 & 0.4 & 0.4 \\
0 & 0 & 0 & 0 & 0 & 0.96 & 0.96 & 0.96 & 0.96 & 0.96 & 0.4 & 0.4 & 0.4 & 0.4 & 0.4 \\
0.96 & 0.96 & 0.96 & 0.96 & 0.96 & 0 & 0 & 0 & 0 & 0 & -0.1 & -0.1 & -0.1 & -0.1 & -0.1 \\
0.96 & 0.96 & 0.96 & 0.96 & 0.96 & 0 & 0 & 0 & 0 & 0 & -0.1 & -0.1 & -0.1 & -0.1 & -0.1 \\
0.96 & 0.96 & 0.96 & 0.96 & 0.96 & 0 & 0 & 0 & 0 & 0 & -0.1 & -0.1 & -0.1 & -0.1 & -0.1 \\
0.96 & 0.96 & 0.96 & 0.96 & 0.96 & 0 & 0 & 0 & 0 & 0 & -0.1 & -0.1 & -0.1 & -0.1 & -0.1 \\
0.4 & 0.4 & 0.4 & 0.4 & 0.4 & -0.1 & -0.1 & -0.1 & -0.1 & -0.1 & 0 & 0 & 0 & 0 & 0 \\
0.4 & 0.4 & 0.4 & 0.4 & 0.4 & -0.1 & -0.1 & -0.1 & -0.1 & -0.1 & 0 & 0 & 0 & 0 & 0 \\
0.4 & 0.4 & 0.4 & 0.4 & 0.4 & -0.1 & -0.1 & -0.1 & -0.1 & -0.1 & 0 & 0 & 0 & 0 & 0 \\
0.4 & 0.4 & 0.4 & 0.4 & 0.4 & -0.1 & -0.1 & -0.1 & -0.1 & -0.1 & 0 & 0 & 0 & 0 & 0 \\
0.4 & 0.4 & 0.4 & 0.4 & 0.4 & -0.1 & -0.1 & -0.1 & -0.1 & -0.1 & 0 & 0 & 0 & 0 & 0
\end{bmatrix}
\end{equation*}

The resulting controlled network is shown in Fig.~\ref{fig:grafo_controllo3_sm} with the clusters $C_1$, $C_2$ and $C_3$ being connected to the bulk with strengths $s_1+w_1=7.8$, $s_2+w_2=5.3$ and $s_3=2.5$, respectively. Consequently, we have that $\lambda_2(\mathcal{L}_1)+s_1+w_1>\lambda_2(\mathcal{L}_2)+s_2+w_2>\lambda_2(\mathcal{L}_3)+s_3+w_3$ and $\lambda_5(\mathcal{L}_2)+s_2+w_2<\lambda_{5}(\mathcal{L}_1)+s_1+w_1<\lambda_5(\mathcal{L}_3)+s_3+w_3$, yielding $\mathcal{I}'=\mathcal{D}'=\{1,2,3\}$. The curves of the synchronization error $\langle \delta_h \rangle_T$ for the three clusters $C_1$, $C_2$ and $C_3$ ($h=\{1,2,3\}$) as function of the coupling strength $d$, shown in Fig.~\ref{fig:errore_sm2}, demonstrate that the controlled multi-agent system displays the predicted synchronization/desynchronization sequence.

\end{document}